\documentclass[a4paper,11pt]{article}
\usepackage{epsfig,epic}
\usepackage{wrapfig}

\newcommand{\resection}[1]{\setcounter{equation}{0}\section{#1}}

\newcommand{\EQ}{\begin{equation}}
\newcommand{\EN}{\end{equation}}
\newcommand{\bea}{\begin{eqnarray}}
\newcommand{\eea}{\end{eqnarray}}

\newcommand{\var}{\varepsilon}
\newcommand{\al}{\alpha}

\newcommand{\goto}{\rightarrow}

\begin{document}
\setcounter{page}{0}
\topmargin 0pt
\oddsidemargin 5mm
\renewcommand{\thefootnote}{\arabic{footnote}}
\newpage
\setcounter{page}{0}
\begin{titlepage}
\begin{flushright}
SISSA 76/2003/FM
\end{flushright}
\vspace{0.5cm}
\begin{center}
{\large {\bf Universal ratios along a line of critical points.}}\\ 
{\large {\bf The Ashkin--Teller model }}\\
\vspace{1.8cm}
{\large Gesualdo Delfino and Paolo Grinza} \\
\vspace{0.5cm}
{\em International School for Advanced Studies (SISSA)}\\
{\em via Beirut 2-4, 34014 Trieste, Italy}\\
{\em INFN sezione di Trieste}\\
{\em E-mail: delfino@sissa.it, grinza@sissa.it}\\
\end{center}
\vspace{1.2cm}

\renewcommand{\thefootnote}{\arabic{footnote}}
\setcounter{footnote}{0}

\begin{abstract}
\noindent
The two-dimensional Ashkin-Teller model provides the simplest example of a 
statistical system exhibiting a line of critical points along which the 
critical exponents vary continously. The scaling limit of both the 
paramagnetic and ferromagnetic phases separated by the critical line are 
described by the sine-Gordon quantum field theory in a given range of its
dimensionless coupling. After computing the relevant matrix elements of 
the order and disorder operators in this integrable field theory, we
determine the universal amplitude ratios along the critical line within the 
two-particle approximation in the form factor approach. 
\end{abstract}

\vspace{.3cm}

\end{titlepage}

\newpage

\resection{Introduction}

The quantitative description of the universality classes of critical behaviour
is the main goal of quantum field theory when applied to statistical 
mechanics. For a statistical system possessing an isolated point of continous
phase transition the canonical characterisation of the scaling behaviour of a 
thermodynamic quantity, say the susceptibility $\chi$, takes the form
\EQ
\lim_{t\rightarrow 0^\pm}\chi=\Gamma_\pm |t|^{-\gamma}\,,
\label{scaling}
\EN
where $t\sim J_c-J$ measures the distance from the critical temperature 
$1/J_c$.
Direct measures or numerical simulations of the system at different 
temperatures will then allow the determination of the critical exponent
$\gamma$ and of the critical amplitudes $\Gamma_\pm$. Contrary to the 
exponent, the amplitudes are not universal, but their ratio $\Gamma_+/\Gamma_-$
is \cite{PHA}. 

In principle, the universal quantities can be computed from the quantum field 
theory encoding the fundamental symmetries of the system. The critical
exponents are yielded by the massless (conformal) field theory describing the
critical point; for the amplitude ratios one needs instead working with the
massive (the mass being an increasing function of $|t|$) field theory 
accounting for the deviations from criticality. 

Consider now the case in which the physical system exhibits a manifold of 
second order phase transition points on which at least some of the critical 
exponents vary continously. To be specific, we will refer to the simplest case 
in which the manifold is a line. Few remarks are in order about the comparison 
of measurements with theoretical preditions in this situation. The off-critical 
Hamiltonian of the system contains now two parameters $J'$ and $J''$ such that 
the critical line corresponds to a curve in the $J'$--$J''$ plane. 
The field theory describing the scaling region close to the critical line will 
depend on a coupling $\beta$ (which has nothing to do with inverse temperature) 
parameterising a line of fixed points of the renormalisation group, as well as 
on a mass scale measuring the distance from criticality. The correspondence 
between the values of $\beta$ and the points of the critical line in the 
$J'$--$J''$ plane is non-universal, i.e. it depends on the microscopic details
of the system like the lattice structure. For each system in the given 
universality class, however, this correspondence can be determined comparing
measurements and field theoretical predictions for at least one of the 
universal quantities (say a critical exponent) which vary continously along 
the line. Once this has been done, field theory yields predictions 
for all other universal quantities in much the same way as in the case of 
isolated critical points.

For a given $\beta$,
the field theory describes a renormalisation group trajectory flowing away
from the point that has been selected on the critical line. Calling $t$ the 
coordinate along this trajectory, equations like (\ref{scaling}) now define 
$\beta$-dependent critical exponents and amplitudes. The trouble is that,
in the generic case, it is not possible to locate the image of the given
trajectory on the $J'$--$J''$ plane. This means that the path along which the
limit has to be taken in (\ref{scaling}) in not known and this kind of 
definitions are practically useless for measuring exponents and 
amplitudes in presence of continously varying critical behaviour.

While the critical exponents can be measured in other ways (e.g. finite size
scaling at criticality), the amplitudes appear essentially out of 
reach in the present case. For the purpose of comparison with the field theory 
predictions, however, we are not interested in the amplitudes themselves,
but rather in their universal combinations. If a duality transformation
relating points on opposite sides of the critical line on the $J'$--$J''$ plane
is available, an amplitude ratio like $\Gamma_+/\Gamma_-$ can in principle be
measured without actually measuring the amplitudes. Its value at a given point 
$P_0$ along the critical line can be determined taking the ratio of the 
susceptibility measured at two dual points close to $P_0$, the theoretical
error going to zero with the distance from $P_0$ on the $J'$--$J''$ plane.

This paper deals with the field theoretical determination of universal
ratios for the simplest class of continously varying critical behaviour.
Critical phenomena in two dimensions are characterised by a number $c$,
called central charge \cite{BPZ}, which increases with the number of degrees 
of freedom of the system. Since for $c<1$ critical exponents are 
only allowed to take discrete values \cite{FQS}, the first possibility for
continously varying exponents opens at $c=1$. This is the central charge of
a free massless boson (Gaussian model), which indeed possesses a continous
one-parameter family of scaling operators. Different statistical models 
exhibiting a line of critical points with continously varying exponents 
renormalise onto the Gaussian model at large distances. These include the 
Ashkin-Teller model \cite{AT} and the 8-vertex model \cite{Baxter}, which
are related by a duality transformation \cite{Wegner}. Their precise relation 
at criticality with the Gaussian model has been determined in the past 
\cite{KB} exploiting the exact Baxter solution of the critical 8-vertex model.

The Ashkin-Teller model is not solved on the lattice away from
criticality\footnote{More precisely, away from its self-dual line.} and the 
universal ratios cannot be computed exactly
apart for the special case in which the model reduces to two decoupled Ising
models. On the field theoretical side, while the relation of the scaling 
limit with sine-Gordon type 
deformations of the Gaussian model has been clear for longtime \cite{Kadanoff},
the quantitative study has been prevented by the need of non-perturbative
methods. Here we exploit the integrability of the sine-Gordon model to
compute the universal ratios along the Ashkin-Teller critical line in the 
two-particle approximation within the form factor approach. 
The form factors of the sine-Gordon model
have been and continue to be the subject of intensive study (see 
\cite{Karowski,Smirnov,Lukyanov,LZ} among other references). The description
of the Ashkin-Teller model, however, relies essentially on the control of 
the order and disorder operators $\sigma$ and $\mu$ which are not among those
considered in these works. We compute all the one- and two-particle matrix
elements needed for our purposes within the framework based on the 
properties of mutual locality between particles and operators.
Duality is used to describe through the sine-Gordon field theory both the
paramagnetic and ferromagnetic phases on the two sides of the critical line.

The paper is organised as follows. We review the phase diagram of the 
Ashkin-Teller model in section~2 and its field theoretical description
in section~3. Section~4 deals with the scattering theory for the scaling
limit around the critical line while section~5 is devoted to form factors.
Correlation functions are discussed in sections~6 and universal ratios are
computed in section~7 before few concluding remarks.

\resection{The isotropic Ashkin-Teller model on the square lattice}

The Ashkin-Teller model \cite{AT,Fan} describes two Ising models coupled 
by a four-spin interaction. The case with the two Ising models having the
same temperature $1/J$ is called `isotropic' and corresponds to the 
Hamiltonian
\EQ
H_{AT}=-\sum_{\langle xy\rangle}\{J[\sigma_1(x)\sigma_1(y)+
                                    \sigma_2(x)\sigma_2(y)]+
J_4\sigma_1(x)\sigma_1(y)\sigma_2(x)\sigma_2(y)\}\,\,,
\label{lattice}
\EN
where $\sigma_1(x)$ and $\sigma_2(x)$ are the two Ising spins at site $x$
and the sum runs over nearest-neighbour pairs $\langle xy\rangle$. 
Each of the transformations $\sigma_1\rightarrow -\sigma_1$, 
$\sigma_2\rightarrow -\sigma_2$ and $\sigma_1\leftrightarrow\sigma_2$ 
leaves the Hamiltonian invariant.

We are interested in the square lattice model for which the transformation
$J\rightarrow -J$ amounts to reversing the spins $\sigma_1$ and $\sigma_2$
on one sublattice. In this case the phase diagram is symmetric under 
reflection about the $J_4$ axis, under which ferromagnetic ordering in 
$\sigma_1$ and $\sigma_2$ becomes antiferromagnetic, and vice versa. With 
this remark in mind we will only refer to the case $J\geq 0$ in the following.

Obviously, the model possesses a critical point in the Ising universality
class at $J=J^*\equiv\frac12\ln(1+\sqrt{2})$ along the decoupling 
line $J_4=0$ (the point marked $D$ in Fig.~1). 
\begin{figure}
\centerline{
\includegraphics[width=13cm]{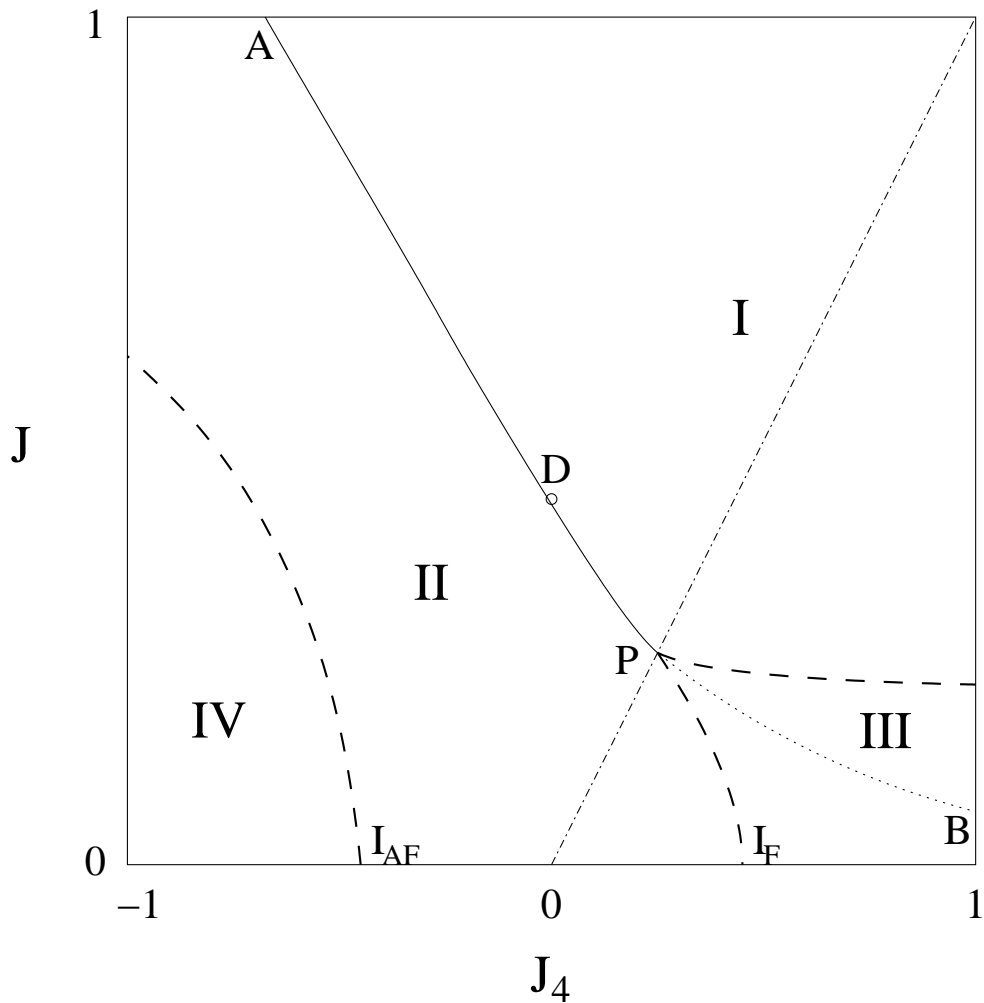}}
\caption{Phase diagram of the isotropic Ashkin--Teller model on the 
square lattice. The self--dual curve is divided into a critical line with
continously varying exponents (continous line) and a non--critical part
(dotted line). The dashed curves are critical lines with Ising critical
exponents. The dash--dotted line is the 4-state Potts model subspace. 
Four different phases are labelled by the roman numerals.}
\end{figure}
The model becomes invariant 
under permutations of the four states $(\sigma_1,\sigma_2)$ along the line
$J=J_4$ which is then the 4-state Potts subspace. Hence, a Potts critical 
point is located at $J=J_4=\frac14\ln 3$ (point $P$ in Fig.~1). The 
antiferromagnetic 4-state Potts model on the square lattice is non-critical
\cite{FS}, so that no other critical point for $J>0$ is implied by the 
analysis of the Potts subspace.

When $J=0$, the Hamiltonian (\ref{lattice}) describes an Ising model in 
the variable $\sigma_1\sigma_2$. Thus the points $I_F$ and $I_{AF}$ in 
Fig.~1 located at $J_4=\pm J^*$ along this line are a ferromagnetic and an
antiferromagnetic Ising critical point, respectively.

As $J_4\rightarrow +\infty$ the energies $\sigma_1(x)\sigma_1(y)$ and 
$\sigma_2(x)\sigma_2(y)$ take the same value, their product being forced to
$1$. Thus in this limit the Hamiltonian (\ref{lattice}) reduces to a single
Ising model with coupling $2J$. A ferromagnetic Ising critical point is 
then located at $(J,J_4)=(J^*/2,+\infty)$.

The model admits a duality transformation \cite{Baxter} which maps the 
Boltzmann weights
\bea
&& \omega_1=e^{2J+J_4+J_0}\nonumber\\
&& \omega_2=e^{-J_4+J_0}\nonumber\\
&& \omega_3=e^{-2J+J_4+J_0}\nonumber
\eea
($J_0$ is an arbitrary constant) into the new ones
\bea
&& \tilde{\omega}_1=\frac12(\omega_1+2\omega_2+\omega_3)\nonumber\\
&& \tilde{\omega}_2=\frac12(\omega_1-\omega_3)\nonumber\\
&& \tilde{\omega}_3=\frac12(\omega_1-2\omega_2+\omega_3)\nonumber\,\,.
\eea
For $J>0$ 
the weights $\tilde{\omega}_j$ are positive (and the corresponding couplings
$\tilde{J}$, $\tilde{J_4}$ and $\tilde{J_0}$ real) provided
\EQ
\cosh 2J>e^{-2J_4}\,\,.
\label{boundary}
\EN
Within the region selected by this condition duality relates points located 
on opposite sides of the self--dual line determined by the equation
\EQ
\sinh 2J=e^{-2J_4}
\label{selfdual}
\EN
(the line A--D--P--B in Fig.~1).

A duality transformation on the $\sigma_1$ spins only maps the square lattice 
Ashkin--Teller model onto a staggered 8-vertex model \cite{Wegner}. In the 
isotropic case (\ref{lattice}) the staggering disappears along the self-dual
line (\ref{selfdual}). Since the unstaggered 8-vertex model is exactly 
solvable, Baxter \cite{Baxter} was able to show that the self-dual line is 
critical for $J_4\leq\frac14\ln 3$ (curve A--D--P in Fig.~1) and non-critical
for $J_4>\frac14\ln 3$ (curve P--B in Fig.~1). The critical exponents vary 
continously along the critical portion of the line. We mention that the 
value $J_4=J^*/2$ selects on the critical line the Fateev-Zamolodchikov 
$Z_4$-parafermionic critical point \cite{FZ}.

No exact lattice results are avalaible for the model away from the self-dual
line. The complete phase diagram, however, has been obtained through a variety
of approximate methods (see \cite{Ditzian}). The critical line with continously
varying critical exponents bifurcates at the Potts point P into two 
critical lines, dual of each other, ending into the previously mentioned 
Ising critical points $(J,J_4)=(J^*/2,+\infty)$ and $I_F$. Another 
critical line originates from the antiferromagnetic Ising critical point 
at $I_{AF}$ and points towards $J_4\rightarrow -\infty$. The exact location
of these three critical lines (dashed in Fig.~1) is unknown. The critical
exponents are expected to stay within the Ising universality class along them.

The four critical lines are the boundaries of four different regions
in the phase diagram of Fig.~1. In I the system is ferromagnetically ordered
with $\langle\sigma_1\rangle$, $\langle\sigma_2\rangle$ and 
$\langle\sigma_1\sigma_2\rangle$ all different from zero; phase II is the 
disordered one in which all these order parameters vanish; phase III exhibits
partial ferromagnetic ordering, with $\langle\sigma_1\rangle=\langle\sigma_2
\rangle=0$ but $\langle\sigma_1\sigma_2\rangle\neq 0$; phase IV is similar
to III but with $\sigma_1\sigma_2$ ordered antiferromagnetically.

\resection{Scaling limit and field theory}

From the field theoretical point of view the critical line with continously
varying exponents of the Ashkin-Teller model must correspond to a line of 
fixed points of the renormalisation group generated by a marginal operator.
The latter can be easily identified looking at the point D on the line where
the model reduces to two non-interacting critical Ising models. The 
four-spin term in (\ref{lattice}) gives in the continuum limit the product
$\varepsilon_1\varepsilon_2$ of the energy densities of the two Ising models.
Since the scaling dimension of the energy operator in the Ising model is $1$,
$\varepsilon_1\varepsilon_2$ is indeed marginal. The existence of a line of 
fixed points implies that this operator remains strictly marginal even away
from the decoupling point and is responsible for the deviation of the 
critical exponents from the Ising values. Accordingly, we write the 
action for the scaling limit around the line of fixed points as
\EQ
{\cal A}_{scaling}={\cal A}^0_1+{\cal A}^0_2+
\tau\,\int d^2x\,(\var_1(x)+\var_2(x))+
\rho\,\int d^2x\,\var_1\var_2(x)\,,
\label{continous}
\EN
where ${\cal A}^0_j$, $j=1,2$ denotes the fixed point Hamiltonian 
of the $j$-th Ising model. The couplings $\tau$ and $\rho$ are non-universal
functions of the lattice couplings $J$ and $J_4$. They determine the distance 
from criticality and the coordinate along the fixed line, respectively.

The Hamiltonian (\ref{continous}) implies 
that the central charge along the line of fixed points is twice 
that of the Ising model, namely $c=1$. This agrees with the fact that
the 8-vertex model along its critical line reduces to the 6-vertex model,
and the latter renormalises al large distances on a free massless boson, 
namely the Gaussian model with central charge equal to $1$. It is well known
\cite{Coleman,Mandelstam} that a Gaussian fixed point in two dimensions also 
admits a description in terms of a Dirac fermion $\psi=\psi_1+i\psi_2$. 
The neutral components $\psi_j$ are the Majorana fermions associated to 
the $j$-th Ising model. The energy operator $\varepsilon_j$ is bilinear in
the fermion ($\varepsilon_j=\psi_j^+\psi_j^-$, with $\psi_j^\pm$ denoting
the two components of the spinor), while the spin operator $\sigma_j$ and 
the disorder operator $\mu_j$ are non-local with respect to it.

The study of the Ashkin-Teller critical line in terms of the Gaussian model
was performed in \cite{KB}.
At the Gaussian fixed point, the boson field can be decomposed into its
holomorphic and antiholomorphic parts as $\varphi(x)=
\phi(z)+\bar{\phi}(\bar{z})$, where we introduced the complex coordinates
$z=x_1+ix_2$ and $\bar{z}=x_1-ix_2$. The scaling operators of the theory
are the vertex operators
\EQ
V_{p,\bar{p}}(x)=e^{i[p\phi(z)+\bar{p}\bar{\phi}(\bar{z})]}\,\,,
\label{vertex}
\EN
with conformal dimensions $(\Delta,\bar{\Delta})=(p^2/{8\pi},
\bar{p}^2/{8\pi})$, scaling dimension $X=\Delta+\bar{\Delta}$ 
and spin $s=\Delta-\bar{\Delta}$. They satisfy the gaussian operator
product expansion
\EQ
V_{p_1,\bar{p}_1}(x)V_{p_2,\bar{p}_2}(0)=
z^{p_1p_2/{4\pi}}\bar{z}^{\bar{p}_1\bar{p}_2/{4\pi}}\,
V_{p_1+p_2,\bar{p}_1+\bar{p}_2}(0)+\ldots\,\,\,.
\label{ope}
\EN
We see from this relation that taking $V_{p_1,\bar{p}_1}(x)$ around
$V_{p_2,\bar{p}_2}(0)$ by sending $z\rightarrow ze^{2i\pi}$ and
$\bar{z}\rightarrow \bar{z}e^{-2i\pi}$ produces a phase factor
$e^{2i\pi\gamma_{1,2}}$, where
\EQ
\gamma_{1,2}=\frac{1}{4\pi}\,(p_1p_2-\bar{p}_1\bar{p}_2)
\label{gamma}
\EN
is called index of mutual locality. If $\gamma_{1,2}$ is an integer the
correlator $\langle V_{p_1,\bar{p}_1}(x)V_{p_2,\bar{p}_2}(0)\rangle$ is
single valued and the two operators are said to be mutually local.
Since $\gamma_{1,1}=2s$, the operators
which are local with respect to themselves (the only ones we are interested
in here) must have integer or half integer spin.

Without loss of generality, we call $\cos\beta\varphi$ the (most relevant
component of the) energy operator ${\cal E}=\varepsilon_1+\varepsilon_2$ which 
drives the system away from criticality. Here $\beta$ is the parameter which
accounts for the continously varying exponents and then parameterises
the critical line; $\beta^2$ is equal to $4\pi$ at the decoupling point where 
the scaling dimension of ${\cal E}$ must be equal to $1$.

All the operators of interest for the description of the
lattice model must be local with respect to the energy operator. 
This locality requirement selects the operators $V_{p,\bar{p}}$ with 
$p-\bar{p}=4\pi m/\beta$, $m$ integer, namely
\EQ
V_{p,p}(x)=e^{ip\varphi(x)}\,
\label{v}
\EN
and
\EQ
U_{n,m}(x)\equiv V_{\frac{n\beta}{2m}+\frac{2\pi}{\beta}m,
                    \frac{n\beta}{2m}-\frac{2\pi}{\beta}m}(x)
=e^{i\left[\frac{n\beta}{2m}\varphi(x)+\frac{2\pi}{\beta}m
\tilde{\varphi}(x)\right]}\,,\hspace{.4cm}n=2s=0,\pm 1,\ldots,
\hspace{.3cm}m=\pm 1,\ldots
\label{u}
\EN
Here we introduced the `dual' boson field $\tilde{\varphi}$ which is
$\phi(z)-\bar{\phi}(\bar{z})$ at criticality and satisfy the relation
\EQ
i\frac{\partial\tilde{\varphi}}{\partial x_a}=\varepsilon_{ab}
\frac{\partial\varphi}{\partial x_b}\,\,.
\label{dual}
\EN              
 
The operators $e^{ip\varphi}$ and $U_{0,m}$ are scalars ($s=0$) and have 
scaling dimensions $X_p=p^2/4\pi$ and $X_{0,m}=\pi m^2/\beta^2$, respectively.
The lowest operators with $|s|=1/2$, i.e. $U_{\pm 1,1}$ and $U_{\pm 1,-1}$
with conformal dimensions $\Delta_{n,m}$ given by
\bea
&& \Delta_{\pm 1,1}=\Delta_{\pm 1,-1}=\frac{1}{8}\left(\frac{\beta^2}{4\pi}
\pm 2+\frac{4\pi}{\beta^2}\right)\,\\
&& \bar{\Delta}_{\pm 1,1}=\bar{\Delta}_{\pm 1,-1}=
\frac{1}{8}\left(\frac{\beta^2}{4\pi}\mp 2+\frac{4\pi}{\beta^2}\right)\,,
\eea         
form the Dirac spinors 
\EQ
\psi=\left(
\begin{array}{c}
\psi_+ \\
{\psi_-} \\
\end{array}
\right)=
\left(
\begin{array}{c}
\psi_1^++i\psi_2^+ \\
{\psi}_1^-+i{\psi}_2^- \\
\end{array}
\right)=
\left(
\begin{array}{l}
U_{1,1} \\
U_{-1,1} \\
\end{array}
\right)
\EN
\EQ
\psi^*=\left(
\begin{array}{c}
\psi^*_+ \\
{\psi}^*_- \\
\end{array}
\right)=
\left(
\begin{array}{c}
\psi_1^+-i\psi_2^+ \\
{\psi}_1^--i{\psi}_2^- \\
\end{array}
\right)=
\left(
\begin{array}{l}
U_{1,-1} \\
U_{-1,-1} \\
\end{array}
\right)\,\,.
\EN 

The Ashkin--Teller operators and their bosonic form \cite{KB} are listed in
the first two columns of Table~1; the scaling dimensions are given in the 
third column. The rest of the table specifies the behaviour of the operators
under the following tranformations:
\begin{center}
\begin{table}[ht]
\hspace{-0.8cm}
\begin{tabular}{|l|c|c|c|c|c|c|c|}\hline
$\Phi$ & Bosonic form & $X_\Phi$ & $Z_2\times Z_2$ & $1\leftrightarrow 2$ & $D_1$
& $D_2$ & $D_\pm$  \\ 
\hline
$\sigma_1$ & & $\frac{1}{8}$ & $-\times +$ & $\sigma_2$ & $\mu_1$ 
& $\sigma_1$ & $\mu_1$ \\
$\sigma_2$ & & $\frac{1}{8}$ & $+\times -$ & $\sigma_1$ & $-\sigma_2$ 
& $\mu_2$ & $\pm\mu_2$ \\
$\mu_1$ & & $\frac{1}{8}$ & $+\times +$ & $\mu_2$ & $\sigma_1$ & $\mu_1$ 
& $\sigma_1$ \\ 
$\mu_2$ & & $\frac{1}{8}$ & $+\times +$ & $\mu_1$ & $\mu_2$ & $\sigma_2$ 
& $\mp\sigma_2$ \\ 
$\psi_1^+$ & $\cos\left(\frac{2\pi}{\beta}\tilde{\varphi}+
\frac\beta2\varphi\right)$ 
& $\frac{\beta^2}{16\pi}+\frac{\pi}{\beta^2}$ 
& $-\times +$ & $\psi_2^+$ & $+$ & $-$ & $-$ \\
$\psi_1^-$ & $\cos\left(\frac{2\pi}{\beta}\tilde{\varphi}
-\frac\beta2\varphi\right)$ 
& $\frac{\beta^2}{16\pi}+\frac{\pi}{\beta^2}$ 
& $-\times +$ & $\psi_2^-$ & $+$ & $+$ & $+$ \\
$\psi_2^+$ & $\sin\left(\frac{2\pi}{\beta}\tilde{\varphi}
+\frac\beta2\varphi\right)$ 
& $\frac{\beta^2}{16\pi}+\frac{\pi}{\beta^2}$ 
& $+\times -$ & $\psi_1^+$ & $-$ & $+$ & $-$ \\
$\psi_2^-$ & $\sin\left(\frac{2\pi}{\beta}\tilde{\varphi}
-\frac\beta2\varphi\right)$ 
& $\frac{\beta^2}{16\pi}+\frac{\pi}{\beta^2}$ 
& $+\times -$ & $\psi_1^-$ & $+$ & $+$ & $+$ \\
${\cal E}=\var_1+\var_2$ & $\cos\beta\varphi$ & $\frac{\beta^2}{4\pi}$ & 
$+\times +$ & + & ${\cal C}$ &${-\cal C}$ & $-$ \\
${\cal C}=\var_1-\var_2$ & $\cos\frac{4\pi}{\beta}\tilde{\varphi}$ & 
$\frac{4\pi}{\beta^2}$ & $+\times +$ & $-$ & ${\cal E}$ &$-{\cal E}$ 
& $-$ \\
${\cal E}_+=\psi_2^+\psi_1^-+\psi_1^+\psi_2^-$ & 
$\sin\frac{4\pi}{\beta}\tilde{\varphi}$ & 
$\frac{4\pi}{\beta^2}$ & $-\times -$ & + & $-{\cal E}_-$ & ${\cal E}_-$ 
& $-$ \\
${\cal E}_-=\psi_2^+\psi_1^--\psi_1^+\psi_2^-$ & 
$\sin\beta\varphi$ & $\frac{\beta^2}{4\pi}$ & $-\times -$ & $-$ & $-{\cal E}_+$
& ${\cal E}_+$ & $-$ \\
${\cal P}=\sigma_1\sigma_2$ & 
$\sin\frac{\beta}{2}\varphi$ & $\frac{\beta^2}{16\pi}$ & $-\times -$ & $-$
& $-\mu_1\sigma_2$ & $\sigma_1\mu_2$ & $\pm\mu_1\mu_2$ \\
${\cal P}^*=\mu_1\mu_2$ & $\cos\frac{\beta}{2}\varphi$ & 
$\frac{\beta^2}{16\pi}$ & $+\times +$ & $+$ & $\sigma_1\mu_2$ & $\mu_1\sigma_2$
& $\mp\sigma_1\sigma_2$ \\
$\sigma_1\mu_2$ & $\cos\frac{2\pi}{\beta}\tilde{\varphi}$ & 
$\frac{\pi}{\beta^2}$ & $-\times +$ & $\mu_1\sigma_2$ & $\mu_1\mu_2$ 
& $\sigma_1\sigma_2$ & $\mp\mu_1\sigma_2$ \\
$\mu_1\sigma_2$ & $\sin\frac{2\pi}{\beta}\tilde{\varphi}$ & 
$\frac{\pi}{\beta^2}$ & $+\times -$ & $\sigma_1\mu_2$ & $-\sigma_1\sigma_2$ 
& $\mu_1\mu_2$ & $\pm\sigma_1\mu_2$ \\ 
${\cal D}_4$ & $\cos 2\beta\varphi$ & $\frac{\beta^2}{\pi}$ & $+\times +$ 
& $+$ & $\tilde{\cal D}_4$ & $\tilde{\cal D}_4$ & $+$ \\ 
$\tilde{\cal D}_4$ & $\cos\frac{8\pi}{\beta}\tilde{\varphi}$ 
& $\frac{16\pi}{\beta^2}$ 
& $+\times +$ & $+$ & ${\cal D}_4$ & ${\cal D}_4$ & $+$ \\ 
$\varepsilon_1\varepsilon_2$ & $ (\partial_a\varphi)^2$ & $2$ 
& $+\times +$ & $+$ 
& $\left(\frac{4\pi}{\beta^2}\partial_a\tilde{\varphi}\right)^2$ 
& $\left(\frac{4\pi}{\beta^2}\partial_a\tilde{\varphi}\right)^2$ 
& $+$ \\ 
\hline
\end{tabular}
\caption{Ashkin-Teller operators with their bosonic counterparts, scaling
dimensions and symmetry properties. When a symmetry transformation sends an
operator into $\pm$ itself, only $\pm$ is indicated.}
\end{table}
\end{center}
\vspace{-0.8cm}
\noindent
{\bf Spin reversal\,.}\hspace{.3cm} We refer to the invariance of the model
under the reversal of all $\sigma_1$ or all $\sigma_2$ spins as 
$Z_2\times Z_2$ symmetry. The transformation
\bea 
\varphi & \rightarrow & -\varphi\nonumber\\
\frac{2\pi}{\beta}\,\tilde{\varphi} & \rightarrow & \pi-\frac{2\pi}{\beta}\,
\tilde{\varphi}\nonumber
\eea
of the bosonic fields is found to correspond to the reversal of the $\sigma_1$ 
spins. The transformation
\bea 
&& \varphi\rightarrow -\varphi\nonumber\\
&& \tilde{\varphi}\rightarrow -\tilde{\varphi}\nonumber
\eea
corresponds instead to the reversal of the $\sigma_2$ spins.

\vspace{.3cm}
\noindent
{\bf Exchange\,.}\hspace{.3cm} The exchange of the two Ising copies
($1\leftrightarrow 2$) is implemented in the bosonic language by the 
transformation
\bea 
\varphi & \rightarrow & -\varphi\nonumber\\
\frac{2\pi}{\beta}\,\tilde{\varphi} & \rightarrow & 
\frac{\pi}{2}-\frac{2\pi}{\beta}\,\tilde{\varphi}\,\,.\nonumber
\eea

\vspace{.3cm}
\noindent
{\bf Semi-duality\,.}\hspace{.3cm} We call semi-duality the transformation
$D_j$ which interchanges $\sigma_j$ and $\mu_j$. We saw in the previous 
section that a transformation of this kind relates the Ashkin--Teller model 
to the 8-vertex model. In the bosonic language $D_1$ corresponds to the 
exchange
\EQ
\frac\beta2\,{\varphi}\leftrightarrow  
-\frac{2\pi}{\beta}\,\tilde{\varphi}\,\,.\nonumber
\EN
and $D_2$ to 
\EQ
\frac\beta2\,{\varphi}\leftrightarrow  
\frac{\pi}{2}-\frac{2\pi}{\beta}\,\tilde{\varphi}\,\nonumber
\EN
Clearly $D_j^2=1$. Notice that $D_1$ is not quite the duality transformation
of the first Ising model. In particular, it changes the sign of $\varepsilon_2$
rather than $\varepsilon_1$. A similar observation applies to $D_2$.

\vspace{.3cm}
\noindent
{\bf Duality\,.}\hspace{.3cm} This is the full duality transformation of 
the Ashkin--Teller model and is obtained composing $D_1$ and $D_2$. Due to 
sign factors there are two possibilities, $D_+=D_1D_2$ and $D_-=D_2D_1$. 
$D_\pm$ correspond to the bosonic transformations
\bea 
\beta\varphi & \rightarrow & \beta\varphi\pm\pi\nonumber\\
\frac{2\pi}{\beta}\,\tilde{\varphi} & \rightarrow & \frac{2\pi}{\beta}\,
\tilde{\varphi}\pm\frac\pi2\,\,.\nonumber
\eea
Notice that $D_+D_-=D_-D_+=1$, $D_\pm^2\neq 1$.

\subsection{The self--dual line}

The field theory describing the self-dual line (\ref{selfdual}) at large 
distances has to be invariant under all the above transformations. 
This requirement selects the action
\EQ
{\cal A}_{sd}=\int d^2x\,\left[\frac{1}{2}\,(\partial_a\varphi)^2
-\sum_{n}\left(g_n\cos 2n\beta\varphi+\tilde{g}_n\cos 8n\frac\pi\beta
\tilde{\varphi}\right)\right]\,,
\label{sd}
\EN
where the $g_n$'s and $\tilde{g}_n$'s are, like $\beta$, non-universal 
functions of $J_4$. This theory is critical (massless) as long as none of the 
operators perturbing the Gaussian term becomes relevant. Since $\cos 2\beta
\varphi$ becomes marginal at $\beta^2=2\pi$ and $\cos 8\pi\tilde{\varphi}/
\beta$ at $\beta^2=8\pi$, the critical range is $2\pi\leq\beta^2\leq 8\pi$.

The $4$-state Potts critical point (P in Fig.~1) is the right end point of 
the Ashkin--Teller critical line. At this point the Ising variables 
$\sigma_1$, $\sigma_2$ and $\sigma_1\sigma_2$ play a completely symmetric
role and must have the same scaling dimension $1/8$. Hence, it follows from
Table~1 that the Potts point corresponds to $\beta^2=2\pi$. 

The relation between $\beta$ and $J_4$ along the critical line for the case 
of the square lattice model with nearest neighbour interactions can be obtained
comparing the energy scaling dimension predicted by the Gaussian theory 
with that coming from the lattice solution. It reads \cite{KB}
\EQ
\frac{4\pi}{\beta^2}=1-\frac2\pi\arcsin\left(\frac{\tanh 2J_4}{\tanh 2J_4-1}
\right)\,,
\label{betaJ}
\EN
from which we see that the limit $J_4\rightarrow-\infty$ corresponds to 
$\beta^2=6\pi$. Thus the Ashkin--Teller critical line with continously 
varying exponents spans only a portion of the critical region of the 
theory (\ref{sd}), namely the range
\EQ
2\pi\leq\beta^2<6\pi\,\,.
\label{range}
\EN

For $\beta^2<2\pi$ at least one of the operators $\cos 2n\beta\varphi$ is 
relevant. The theory (\ref{sd}) is massive and corresponds to the non-critical 
part of the self-dual line. The relation (\ref{betaJ}) does not hold in this
region.

\subsection{Breaking duality}

The field theory describing the model on the two sides of the self--dual 
line is obtained adding to the self--dual
action (\ref{sd}) the operators $\cos(2n-1)\beta\varphi$, 
which preserve the spin reversal and exchange symmetries but are odd under 
duality
\EQ
{\cal A}_{AT}={\cal A}_{sd}-\sum_{n}\tau_n\int d^2x\,\cos(2n-1)\beta\varphi
\,,
\label{AT}
\EN
where the $\tau_n$'s are functions of the lattice couplings $J$ and $J_4$. 

To describe the scaling regions around the critical line with continously
varying exponents we keep the only operator which is relevant in the range
(\ref{range}). Thus we are left with the euclidean sine-Gordon action (we set
$\tau_1=\tau$) 
\EQ
{\cal A}_{SG}=\int d^2x\,\left(\frac{1}{2}\,(\partial_a\varphi)^2-
\tau\cos\beta\varphi\right)
\label{sg}
\EN
which is the bosonic version of (\ref{continous}). 

For $\tau>0$ this action describes the scaling region of the paramagnetic
phase II in which the vacuum $|0\rangle$ located at $\varphi=0$ is invariant 
under spin reversal and exchange symmetry and we have 
\bea
&& \langle 0|\sigma_1|0\rangle=\langle 0|\sigma_2|0\rangle=0\nonumber\\
&& \langle 0|\sigma_1\sigma_2|0\rangle=
   \langle 0|\sin\frac{\beta\varphi}{2}|0\rangle=0\,\,.\nonumber
\eea

For $\tau<0$ the two vacua $|0_\pm\rangle$ located at $\beta\varphi=\pm\pi$
are the image of the high--temperature vacuum $|0\rangle$ through the duality
transformations $D_\pm$. 
The vacua $|0_+\rangle$ and $|0_-\rangle$ are interchanged by the spin
reversal and exchange transformations, so that the internal symmetries of the
model are spontaneously broken (ferromagnetic phase I). We have
\bea
&& \langle 0_+|\sigma_1|0_+\rangle=\langle 0_+|\sigma_2|0_+\rangle=
   \pm M_\sigma\nonumber \\
&& \langle 0_-|\sigma_1|0_-\rangle=-\langle 0_-|\sigma_2|0_-\rangle=
   \pm M_\sigma\nonumber \\
&& \langle 0_\pm|\sigma_1\sigma_2|0_\pm\rangle=\langle 
0_\pm|\sin\frac{\beta\varphi}{2}|0_\pm\rangle=\pm M_{\cal P}\,\nonumber
\eea
where $M_\sigma$ and $M_{\cal P}$ are positive. The last equation follows 
from the relations
\bea
&& \langle 0_+|e^{i\alpha\varphi}|0_+\rangle=
   \langle 0_-|e^{-i\alpha\varphi}|0_-\rangle\nonumber\\
&& \langle 0_\pm|e^{i\alpha\varphi}|0_\pm\rangle=e^{\pm2i\pi\alpha/\beta}
   \langle 0_\pm|e^{-i\alpha\varphi}|0_\pm\rangle\,\,.\nonumber
\eea

\subsection{Bifurcation at the Potts critical point}

For $\beta^2<2\pi$ more operators become relevant in the action (\ref{AT}).
Neglecting all irrelevant terms in the range $8\pi/9<\beta^2<2\pi$ leads to
the double sine-Gordon action ($g_1\equiv g$) \cite{Kadanoff}
\EQ
{\cal A}_{DSG}=\int d^2x\,\left(\frac{1}{2}\,(\partial_a\varphi)^2-
\tau\cos\beta\varphi-g\cos 2\beta\varphi\right)\,\,.
\label{dsg}
\EN
This quantum field theory has been analysed in Ref.~\cite{msg}.  
The mechanism through which it accounts for the bifurcation of the critical 
line at the Potts critical point is easily understood already at the 
classical level. 
To see this let us fix a value of $\beta$ in the considered range and treat 
for a moment both $\tau$ and $g$ as free parameters. For $g=0$ and fixed 
$\tau>0$ the vacuum of the theory (\ref{dsg}) (i.e. the minimum of the 
potential invariant under the symmetries of the paramagnetic phase) is located 
at $\varphi=0$. It stays there as we decrease $g$ down to a critical value 
$g_c$, classically equal to $-\tau/4$, where the quadratic term of the 
potential vanishes. For $g<g_c$ we have a maximum at $\varphi=0$ and two
new minima located symmetrically with respect to it. Thus an Ising phase 
transition occurred at $g_c$. 
\begin{figure}
\centerline{
\includegraphics[width=10cm]{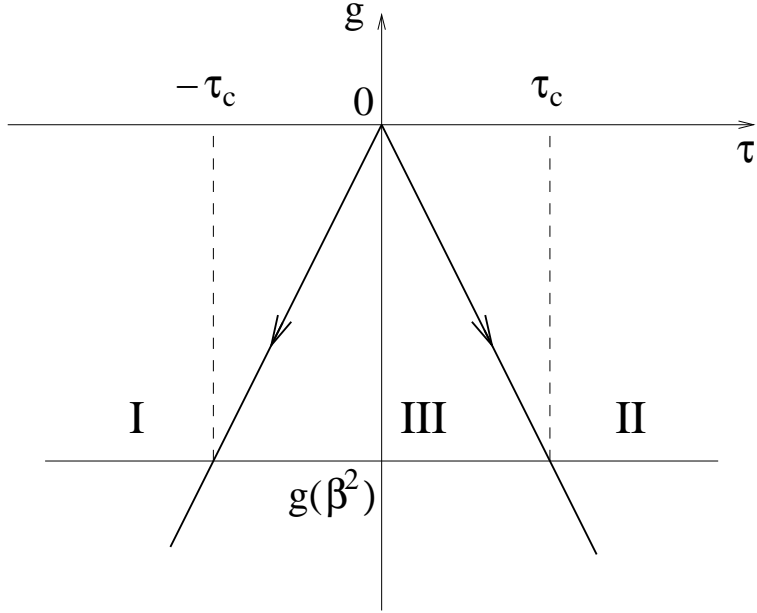}}
\caption{Schematic phase diagram of the double sine-Gordon quantum
field theory (\ref{dsg}) for fixed $\beta^2<2\pi$. Two massless trajectories
starting from the Gaussian fixed point at the origin flow towards infrared
Ising fixed points and divide the plane into three different phases. The 
points at $\pm\tau_c$ on these trajectories correspond to dual points along
the Ising critical lines bifurcating from point P in Fig.~1.}
\end{figure}
Similar considerations can be repeated for $\tau<0$ (starting with one of the 
vacua at $\varphi=\pm\pi$ for $g=0$) and the resulting 
picture can be confirmed at the quantum level \cite{msg}. 
In summary, the lower $\tau$-$g$ half plane is divided into
three regions (Fig.~2) by two massless trajectories (corresponding classically 
to $g=\pm\tau/4$ but whose precise location is unknown in the quantum theory) 
along which a flow from the Gaussian fixed point at the origin to 
infrared Ising fixed points takes place. Region II is a 
paramagnetic phase where the vacuum is located at $\varphi=0$; region I is 
a ferromagnetic phase dual to II where $\langle\beta\varphi\rangle$ equals
$\pi$ or $-\pi$; in region III $\langle\beta\varphi\rangle$ interpolates
smoothly from $0$ to $\pm\pi$ taking the value $\pm\pi/2$ at $\tau=0$. In this 
latter phase we have $\langle\sigma_1\sigma_2\rangle\neq 0$; it is less clear
to us how to argue that $\langle\sigma_1\rangle=\langle\sigma_2\rangle=0$.

In order to make contact with the phase diagram of Fig.~1 we need to recall
that $g$ is not a free parameter at fixed $\beta$. Indeed, both $\beta$ and 
$g$ are determined by the value that $J_4$ takes along the self--dual line.
It is the value $g(\beta^2)$ that determines the distance $\tau_c$ of the 
Ising transition points from the given point along the self--dual line (see 
Fig.~2). The  
phase diagram requires that $g(\beta^2)$ vanishes at $\beta^2=2\pi$ and then
decreases with $\beta^2$. The result in the coupling space of the theory
(\ref{dsg}) is shown in Fig.~3.
\begin{figure}
\centerline{
\includegraphics[width=8cm]{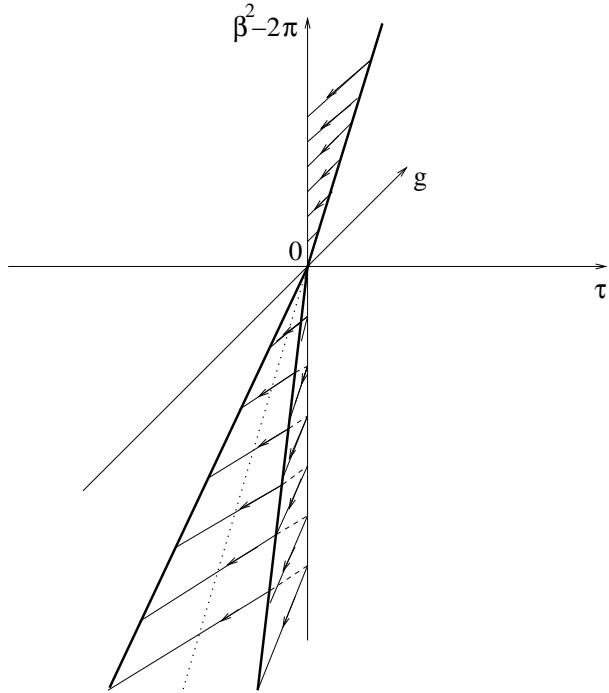}}
\caption{Like in Fig.~2 but with varying $\beta^2$. The oriented 
lines indicate the flow towards larger distances along the critical surfaces.
The thick lines are determined by the intersection of these critical surfaces
with the image of the $J$-$J_4$ plane of Fig.~1.}  
\end{figure}
For $\pi/2<\beta^2<2\pi$ the non--critical part of the self--dual line is 
described by the action (\ref{dsg}) with $\tau=0$. As the Potts critical 
point is approached from below along the line ($\beta^2\rightarrow 2\pi^-$,
$g\rightarrow 0^-$) the operator $\cos 2\beta\varphi$
becomes marginally relevant implying an exponential rather than power law 
divergence of the correlation length. The same massive field theory describes 
the $q\rightarrow 4^+$ limit of the $q$-state Potts model at $T=T_c$ \cite{q4}.

\resection{Scattering theory}

According to the discussion of the previous section the scaling limit of 
the Ashkin--Teller model around the critical line with continously varying 
exponents is described by the sine-Gordon theory (\ref{sg}). This 
quantum field theory is integrable and the associated elastic and factorised
scattering matrix is exactly known (\cite{ZZ} and references therein).

The elementary excitations are the soliton $A_+$ and antisoliton $A_-$ 
which interpolate between adjacent vacua of the periodic pontential. While
being topologic excitations of the bosonic action (\ref{sg}) they correspond
to the fermions $\psi$ and $\psi^*$ of the equivalent fermionic model (the
massive Thirring model) \cite{Coleman,Mandelstam}. Actually, the integer $m$
in (\ref{u}) measures precisely the topologic charge and all
the operators with $m=1$ ($-1$) create a soliton (antisoliton) when acting on 
the vacuum of the theory. Writing $A_\pm=(A_1\pm iA_2)/\sqrt{2}$, the 
operators $\psi_i$ and $\sigma_i\mu_{i+1\,(\mbox{mod}\,\,2)}$  are both 
suitable interpolating operators for the neutral component $A_i$. To be
definite, we will refer to the latter choice in the following (see Table~2).
\begin{figure}
\centerline{
\includegraphics[width=5cm]{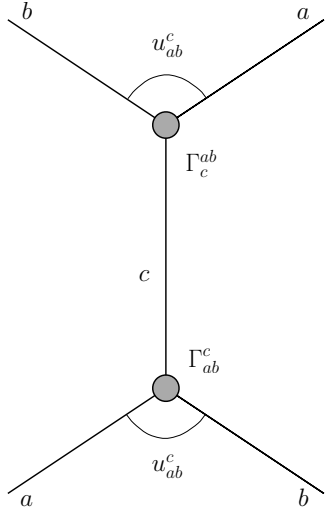}}
\caption{Simple pole diagram associated to Eq.~(\ref{gabc}).}
\end{figure}
The scattering of the particles $A_1$ and $A_2$ in the integrable theory is 
completely determined by the relations\footnote{The on-shell energy and 
momentum of a particle of species $a$ are parameterised as 
$(p^0,p^1)=(m_a\cosh\theta,m_a\sinh\theta)$.} 
\EQ
A_i(\theta_1)A_j(\theta_2)=\sum_{k,l=1,2}S_{A_iA_j}^{A_kA_l}(\theta_1-\theta_2)
A_l(\theta_2)A_k(\theta_1)\,\,
\label{FZ}
\EN
with the non-zero scattering amplitudes given by \cite{ZZ}
\bea
&& S_{A_1A_1}^{A_1A_1}(\theta)=S_{A_2A_2}^{A_2A_2}(\theta)=\frac{S(\theta)+
S_+(\theta)}{2}
\nonumber\\
&& S_{A_1A_1}^{A_2A_2}(\theta)=S_{A_2A_2}^{A_1A_1}(\theta)=\frac{S_+(\theta)-
S(\theta)}{2}
\nonumber\\
&& S_{A_1A_2}^{A_1A_2}(\theta)=S_{A_2A_1}^{A_2A_1}(\theta)=\frac{S(\theta)+
S_-(\theta)}{2}
\nonumber\\
&& S_{A_1A_2}^{A_2A_1}(\theta)=S_{A_2A_1}^{A_1A_2}(\theta)=\frac{S(\theta)-
S_-(\theta)}{2}
\,\,,
\eea
where
\bea
&& S_+(\theta)=-\frac{\sinh\frac{\pi}{2\xi}(\theta+i\pi)}
               {\sinh\frac{\pi}{2\xi}(\theta-i\pi)}S(\theta)\nonumber \\
&& S_-(\theta)=-\frac{\cosh\frac{\pi}{2\xi}(\theta+i\pi)}
               {\cosh\frac{\pi}{2\xi}(\theta-i\pi)}S(\theta)\nonumber \\
&& S(\theta)=-\exp\left\{-i\int_0^\infty\frac{dx}{x}
\frac{\sinh\frac{x}{2}\left(1-
\frac{\xi}{\pi}\right)}{\sinh\frac{x\xi}{2\pi}\cosh\frac{x}{2}}
\sin\frac{\theta x}{\pi}\right\}\nonumber \\
&& \xi=\frac{\pi\beta^2}{8\pi-\beta^2}\,\,.\nonumber
\eea

For $\beta^2<4\pi$ (attractive regime) the above amplitudes possess simple 
poles in the physical strip $\mbox{Im}\theta\in(0,\pi)$ corresponding to
bound states (breathers) $B_n$. The poles are located at $\theta=i(\pi-n\xi)$
($n$ even for $S_+(\theta)$ and odd for $S_-(\theta)$) and determine the 
masses of the breathers 
\EQ
m_n=2M\sin\frac{n\xi}{2}\,,\hspace{1cm}1\leq n<\left[\frac\pi\xi\right]
\label{mn}
\EN
where $M$ is the mass of the particles $A_j$ and $[x]$ denotes the integer
part of $x$. The lightest breather $B_1$ is the particle interpolated by the
boson $\varphi$ and then is odd under all the internal symmetries of the 
model. The breather $B_n$ can be seen as a bound state of $n$ breathers $B_1$,
so that its symmetry properties are those summarised in Table~2.

\begin{table}[ht]
\begin{center}
\begin{tabular}{|l|c|c|c|}\hline
Particle & Creating & $Z_2\times Z_2$ & $1\leftrightarrow 2 $ \\ 
         & operator &                 &                       \\ \hline
$A_1$ & $\sigma_1\mu_2$ & $-\times +$ & $A_2$ \\
$A_2$ & $\mu_1\sigma_2$ & $+\times -$ & $A_1$ \\
$B_{2k+1}$ & $\sin\beta\varphi$ & $-\times -$ & $-$ \\
$B_{2k}$   & $\cos\beta\varphi$ & $+\times +$ & $+$ \\ 
\hline
\end{tabular}
\end{center}
\caption{Particles, intepolating operators and their symmetries.}
\end{table}

The scattering of the breathers with the elementary excitations and with
themselves is completely diagonal (initial and final states are identical).
The corresponding amplitudes are
\EQ
S_{A_jB_n}(\theta)=t_{\frac12+\frac{n\xi}{2\pi}}(\theta)\times
\left\{
\begin{array}{l}
\prod_{j=1}^{(n-1)/2}t^2_{\frac12+(n-2j)\frac{\xi}{2\pi}}(\theta)\,,
\hspace{.5cm}n\,\,\,\mbox{odd} \\ \\
t_\frac12(\theta)
\prod_{j=1}^{(n-2)/2}t^2_{\frac12+(n-2j)\frac{\xi}{2\pi}}(\theta)\,,
\hspace{.5cm}n\,\,\,\mbox{even} \\ 
\end{array}
\right.
\EN
\EQ
S_{B_mB_n}(\theta)=t_{(m+n)\frac{\xi}{2\pi}}(\theta)
                   t_{1-|m-n|\frac{\xi}{2\pi}}(\theta)
\prod_{j=1}^{\mbox{min}\,(m,n)-1}t^2_{(|m-n|+2j)\frac{\xi}{2\pi}}(\theta)
\label{ab}
\EN
where
\EQ
t_\alpha(\theta)=\frac{\tanh\frac12(\theta+i\pi\alpha)}
                      {\tanh\frac12(\theta-i\pi\alpha)}\,\,.
\EN
The simple pole at $\theta=i(\pi+n\xi)/2$ in the amplitude $S_{A_jB_n}(\theta)$
corresponds to the appearance as a bound state in the channel $A_jB_n$ of the 
particle $A_{j+1\,(\mbox{mod}\,2)}$ ($n$ odd) or $A_j$ ($n$ even). 
The simple pole at $\theta=i(n+m)\xi/2$ ($\theta=i[\pi-|n-m|\xi/2]$) in the 
amplitude $S_{B_mB_n}(\theta)$ signals that $B_{n+m}$ ($B_{|n-m|}$) appears as 
a bound state in the $B_mB_n$ channel. 
\begin{figure}
\centerline{
\includegraphics[width=9cm]{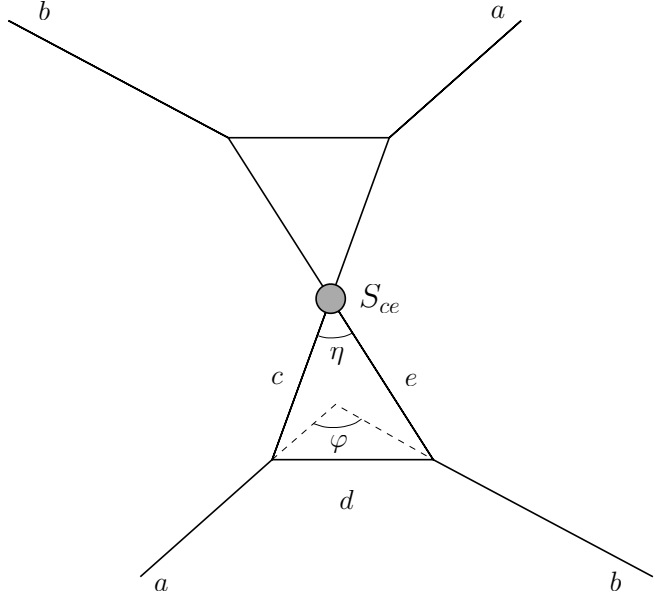}
}
\caption{Double pole diagram associated to Eq.~(\ref{double}).}
\end{figure}
A crossed channel pole located at $i\pi-\theta$ is associated to each of 
these direct channel poles. A direct channel pole at $\theta=iu_{ab}^c$ 
corresponding to the particle $c$ appearing
as bound state in the $ab$ channel allows the determination of the 
three-particle coupling $\Gamma_{ab}^c$ through the relation (see Fig.~4)
\EQ
S_{ab}(\theta\simeq iu_{ab}^c)\simeq 
i\frac{\Gamma_{ab}^c\Gamma_c^{ab}}{\theta-iu_{ab}^c}\,\,
\label{gabc}
\EN
where $\Gamma_c^{ab}=C_aC_bC_c\Gamma_{ab}^c$, $\Gamma_{ab}^c=
C_aC_c\Gamma_{bc}^a$ with $C_{A_j}=1$, $C_{B_n}=(-1)^n$. We also note the 
property $\Gamma_{A_iA_j}^{B_n}=(-1)^{i+j}\Gamma_{A_jA_i}^{B_n}$.

The second order poles do not correspond to
bound states and can be explained in terms of multiscattering processes
of the type shown in Fig.~5 \cite{CT,Goebel}. In the vicinity of such a pole 
located at $\theta=i\varphi$ the scattering amplitude can be written as
\EQ
S_{ab}(\theta\simeq i\varphi)\simeq 
\frac{\Gamma_{ad}^c\Gamma_{db}^e\Gamma^{ad}_c\Gamma^{db}_e 
S_{ce}(i\eta)}{(\theta-i\varphi)^2}\,\,.
\label{double}
\EN

\resection{Form factors}

The knowledge of the scattering amplitudes allows the determination of the 
form factors 
\EQ
F^\Phi_{a_1\ldots a_n}(\theta_1,\ldots,\theta_n)=
\langle 0|\Phi(0)|a_1(\theta_1)\ldots a_n(\theta_n)\rangle\,\,.
\label{fn}
\EN

Many results are known about the form factors of the sine-Gordon model
\cite{Karowski,Smirnov,Lukyanov,LZ}. Here we will recall how the one- and 
two-particle 
form factors can be computed within the general approach. These matrix 
elements for the order and disorder operators $\sigma_j$ and $\mu_j$, which 
are crucial for the description of the Ashkin--Teller model, were first 
considered in \cite{previous}. 

We use for the two-particle form factors of a scalar operator $\Phi(x)$ the
notation
\EQ
F^\Phi_{ab}(\theta_1-\theta_2)=\langle 0|\Phi(0)|a(\theta_1)b(\theta_2)\rangle\,,
\label{ff}
\EN
the dependence on the rapidity difference being a consequence of Lorentz 
invariance.
The matrix elements satisfy the equations \cite{Karowski,Smirnov,YZ}
\bea
&& F^\Phi_{ab}(\theta)=S_{ab}^{cd}(\theta)F^\Phi_{dc}(-\theta)
\label{unitarity}\\
&& F^\Phi_{ab}(\theta+2i\pi)=l_{\Phi,a}F^\Phi_{ba}(-\theta)
\label{crossing}\\
&& \mbox{Res}_{\theta=i\pi}F^\Phi_{\bar{a}b}(\theta)=i\delta_{ab}\,
(1-l_{\Phi,a})\langle\Phi\rangle
\label{kinetic}\\
&& \mbox{Res}_{\theta=iu_{ab}^c}F^\Phi_{ab}(\theta)=i\Gamma_{ab}^c\,F_c^\Phi
\label{bound}
\eea
where the factor $l_{\Phi,a}$ takes into account the mutual locality between 
$\Phi$ and the particle $a$ (see Table~3). If $\phi_a$ is the operator
which interpolates the particle $a$, then $l_{\Phi,a}=
e^{2i\pi\gamma_{\Phi,\phi_a}}$. For the operators having a bosonic expression,
$\gamma_{\Phi,\phi_a}$ is given by (\ref{gamma}). For $\sigma_j$ and $\mu_j$
the results of Table~3 follow from the properties $l_{\sigma_j,\sigma_k}=
l_{\mu_j,\mu_k}=1$ and $l_{\sigma_j,\mu_k}=(-1)^{\delta_{jk}}$ (notice that
the even breathers appear in the $A_jA_j$ channel while the odd ones appear
in the $A_1A_2$ channel).
\begin{table}[ht]
\begin{center}
\begin{tabular}{|l|c|l|l|}\hline
$\Phi$ & Bosonic & $l_{\Phi,A_j}$ & $l_{\Phi,B_n}$ \\ 
       & form    &                     &            \\ \hline
$\sigma_i$ & & $(-1)^{\delta_{ij}+1}$ & $(-1)^n$ \\
$\mu_i$    & & $(-1)^{\delta_{ij}}$   & $(-1)^n$ \\
${\cal E}=\var_1+\var_2$ & $\cos\beta\varphi$ & $+1$ & $+1$ \\
${\cal C}=\var_1-\var_2$ & $\cos\frac{4\pi}{\beta}\tilde{\varphi}$ 
& $+1$ & $+1$ \\
${\cal E}_+=\bar{\psi}_1\psi_2+\bar{\psi}_2\psi_1$ & 
$\sin\frac{4\pi}{\beta}\tilde{\varphi}$ & $+1$ & $+1$  \\
${\cal E}_-=\bar{\psi}_1\psi_2-\bar{\psi}_2\psi_1$ & $\sin\beta\varphi$ 
& $+1$ & $+1$ \\
${\cal P}=\sigma_1\sigma_2$ & $\sin\frac{\beta}{2}\varphi$ 
& $-1$ & $+1$ \\
${\cal P}^*=\mu_1\mu_2$ & $\cos\frac{\beta}{2}\varphi$ & $-1$ & $+1$ \\
$\sigma_1\mu_2$ & $\cos\frac{2\pi}{\beta}\tilde{\varphi}$ & $+1$ & $+1$ \\
$\mu_1\sigma_2$ & $\sin\frac{2\pi}{\beta}\tilde{\varphi}$ & $+1$ & $+1$ \\ 
\hline
\end{tabular}
\caption{Factors of mutual locality between particles and operators
entering Eqs.~(\ref{crossing}) and (\ref{kinetic}).}
\end{center}
\end{table}

The form factors are further constrained by the asymptotic bound \cite{immf}
\EQ
\lim_{|\theta|\rightarrow\infty}F^\Phi_{ab}(\theta)\sim e^{y_\Phi|\theta|}
\EN
\EQ
y_\Phi\leq \frac{X_\Phi}{2}\,,
\label{asymp}
\EN
where $X_\Phi$ is the operator scaling dimension.
The non-zero two-particle matrix elements on the elementary excitations $A_1$ 
and $A_2$ in the disordered phase are then uniquely determined to be 
\bea
&& F^{\mu_l}_{A_jA_j}(\theta)=\frac{i\pi\langle\mu\rangle}{2\xi\omega(i\pi)}\,
\frac{F_0(\theta)}{\sinh\frac{\pi}{2\xi}(\theta-i\pi)}
\left[\omega(\theta)+(-1)^{l+j}\omega(2i\pi-\theta)\right]\\
&& F^{\cal E}_{A_jA_j}(\theta)=ic_1\,
\frac{\cosh\frac{\theta}{2}}{\sinh\frac{\pi}{2\xi}
(\theta-i\pi)}\,F_0(\theta) \label{energy}\\
&& F^{\cal C}_{A_jA_j}(\theta)=c_2\,(-1)^j\,F_0(\theta)\,\,\\
&& F^{{\cal E}_+}_{A_1A_2}(\theta)=F^{{\cal E}_+}_{A_2A_1}(\theta)=
ic_2\,F_0(\theta)\,\,\\
&& F^{{\cal E}_-}_{A_1A_2}(\theta)=-F^{{\cal E}_-}_{A_2A_1}(\theta)=
-c_1\,\frac{\cosh\frac{\theta}{2}}
{\cosh\frac{\pi}{2\xi}(\theta-i\pi)}\,F_0(\theta)\,\,\\
&& F^{{\cal P}}_{A_1A_2}(\theta)=-F^{{\cal P}}_{A_2A_1}(\theta)=
-\frac{\pi}{\xi}\langle{\cal P}^*\rangle\,
\frac{F_0(\theta)}{\cosh\frac{\pi}{2\xi}(\theta-i\pi)}\,\,\\
&& F^{{\cal P}^*}_{A_jA_j}(\theta)=\frac{i\pi}{\xi}\langle{\cal P}^*\rangle\,
\frac{F_0(\theta)}{\sinh\frac{\pi}{2\xi}(\theta-i\pi)}\,\,.
\eea
Here $c_1$ and $c_2$ are normalisation constants, $\langle\mu\rangle\equiv
\langle\mu_1\rangle=\langle\mu_2\rangle$, and the functions
\EQ
\omega(\theta)=\exp\left\{2\int_0^\infty\frac{dx}{x}\,
\frac{\sinh\left(1-\frac{\xi}{\pi}\right)x}
{\sinh\frac{x\xi}{\pi}}\,\frac{\sin^2\frac{\theta x}{2\pi}}{\sinh 2x}\right\}
\EN
\EQ
F_0(\theta)=-i\sinh\frac{\theta}{2}\,
\exp\left\{\int_0^\infty\frac{dx}{x}\,\frac{\sinh\frac{x}{2}\left(1-
\frac{\xi}{\pi}\right)}{\sinh\frac{x\xi}{2\pi}\,\cosh\frac{x}{2}}\,
\frac{\sin^2\frac{(i\pi-\theta)x}{2\pi}}{\sinh x}\right\}\,\,
\EN
satisfy the equations
\EQ
\omega(\theta)=\omega(-\theta)\,\,,\hspace{1cm}
\omega(\theta+2i\pi)=-\frac{\sinh\frac{\pi}{2\xi}(\theta+i\pi)}
{\sinh\frac{\pi}{2\xi}(\theta-i\pi)}\omega(\theta-2i\pi)
\EN
\EQ
F_0(\theta)=S(\theta)F_0(-\theta)\,\,,\hspace{1cm}F_0(\theta+2i\pi)=
F_0(-\theta)\,\,.
\EN
For large values of $|\theta|$ they behave as 
\bea
&& \omega(\theta)\sim\exp\left[\left(\frac{\pi}{\xi}-1\right)\frac{|\theta|}{4}
\right] \\ 
&& F_0(\theta)\sim\exp\left[\left(\frac{\pi}{\xi}+1\right)\frac{|\theta|}{4}
\right]
\,\,.
\eea
For $\xi<\pi/2$, the analytic continuation
\bea
\omega(2i\pi-\theta) &=& -\frac{\sinh[(\theta-i(\pi-2\xi))/4]
                           \sinh[(\theta+i(\pi-2\xi))/4]}
{\cos^2[(\pi-2\xi)/4]}\nonumber\\
&\times &\exp\left\{2\int_0^\infty\frac{dx}{x}\,
\frac{\sinh\left(1-\frac{3\xi}{\pi}\right)x}
{\sinh\frac{x\xi}{\pi}\sinh 2x}\,\sin^2\frac{(2i\pi-\theta)x}{2\pi}
\right\}
\eea
is convergent for real rapidity values.

The breather-breather form factors can be 
written in the form
\EQ
F^\Phi_{B_nB_m}(\theta)=P^{\Phi}_{B_nB_m}(\theta)
\left(\cosh\frac\theta2\right)^{\frac12(1-l_{\Phi,B_n})(-1)^{\delta{n,m}}}
\frac{F^{min}_{B_nB_m}(\theta)}{D_{B_nB_m}(\theta)}\,\,.
\label{param}
\EN
Here
\EQ
F^{min}_{B_mB_n}(\theta)=
T_{(m+n)\frac{\xi}{2\pi}}(\theta)T_{1-|m-n|\frac{\xi}{2\pi}}(\theta)
\prod_{j=1}^{\mbox{min}\,(m,n)-1}T^2_{(|m-n|+2j)\frac{\xi}{2\pi}}(\theta)\,\,.
\EN
The function
\EQ
T_{\al}(\theta)=\exp\left\{2\int_0^\infty\frac{dt}{t}\frac{\cosh\left(
\al - \frac{1}{2}\right)t}{\cosh\frac{t}{2}\sinh
t}\sin^2\frac{(i\pi-\theta)t}{2\pi}\right\}
\EN
solves the equations
\EQ
T_{\alpha}(\theta)=-t_\alpha(\theta)T_\alpha(-\theta)\,,\hspace{1cm}
T_{\alpha}(\theta+2i\pi)=T_\alpha(-\theta)
\EN
and behaves asymptotically as
\EQ
T_{\al}(\theta)\sim\exp(|\theta|/2)\,, \hspace{1cm}|\theta|\goto\infty \,\,.
\EN
In particular
\EQ
T_0(\theta)=T_1(\theta)=-i\sinh\frac\theta2\,\,.
\EN

The denominator $D_{B_nB_m}(\theta)$ in (\ref{param}) accounts for the pole 
structure of the form factors and can be written as (see \cite{immf})
\EQ
D_{B_mB_n}(\theta)={\cal P}_{(m+n)\frac{\xi}{2\pi}}(\theta)
                   {\cal P}_{1-|m-n|\frac{\xi}{2\pi}}(\theta)
\prod_{j=1}^{\mbox{min}\,(m,n)-1}{\cal W}_{(|m-n|+2j)\frac{\xi}{2\pi}}(\theta)
\EN
with
\EQ
{\cal P}_{\al}(\theta)=\frac{\cos\pi\al-\cosh\theta}{2\cos^2\frac{\pi\al}{2}}\,,\hspace{.5cm}\alpha\neq 1
\EN
\EQ
{\cal P}_1(\theta)=1
\EN
\EQ
{\cal W}_{\alpha}(\theta)={\cal P}_{\al}(\theta){\cal P}_{1-\al}(\theta)\,\,.
\EN

The last ingredient of (\ref{param}) are the polynomials
\EQ
P^\Phi_{ab}(\theta)=\sum_{k=0}^{N^\Phi_{ab}} c^{\Phi;k}_{ab}\,(\cosh\theta)^k
\EN
whose total degree is constrained by the the asymptotic bound
(\ref{asymp}) and whose coefficients are determined by the residue equations
(\ref{kinetic}) and (\ref{bound}) together\footnote{See also (\ref{zero}) for
the energy operator ${\cal E}(x)$.} with the relation \cite{immf}
\EQ
F^\Phi_{ab}(\theta\simeq i\varphi)\simeq 
i\frac{\Gamma_{ad}^c\Gamma_{db}^eF^\Phi_{ce}(i\eta)}{\theta-i\varphi}
\label{doubleff}
\EN
associated to the double poles (\ref{double}) in the scattering amplitudes.
In particular, one finds $N_{B_1B_1}^\Phi=0$ and $N_{B_2B_2}^\Phi=1$ for 
$\Phi=\mu,\,{\cal E},\,{\cal P}^*$, and $N_{B_1B_2}^{\cal P}=1$.

Concerning the channel $A_jB_n$, the property
\EQ
F^\Phi_{A_jB_n}(\theta)=(-1)^nF^\Phi_{B_nA_j}(\theta)
\EN
has to be taken into account. We find that the minus sign appearing 
for odd breathers (and implying $\Gamma_{A_jB_{2k+1}}^{A_{j+1\,(mod\,2)}}=
-\Gamma_{B_{2k+1}A_j}^{A_{j+1\,(mod\,\,2)}}$) is needed if solutions 
compatible with the asymptotic bound (\ref{asymp}) are to be found for the
operators $\sigma_i$ and $\sigma_i\mu_{i+1\,(mod\,2)}$. 
Then one can write
\EQ
F^\Phi_{A_jB_n}(\theta)=P^{\Phi}_{A_jB_n}(\theta)
\left(\cosh\frac\theta2\right)^{\delta_{n,odd}\delta_{1,l_{\Phi,B_n}}}
\frac{F^{min}_{A_jB_n}(\theta)}{D_{A_jB_n}(\theta)}\,,
\label{param2}
\EN
where
\EQ
F^{min}_{A_jB_n}(\theta)=
\prod_{j=0}^{n-1}T_{\frac12+(n-2j)\frac{\xi}{2\pi}}(\theta)
\EN
\EQ
D_{A_jB_n}(\theta)={\cal P}_{\frac12+\frac{n\xi}{2\pi}}(\theta)\times
\left\{
\begin{array}{l}
\prod_{j=1}^{(n-1)/2}{\cal W}_{\frac12+(n-2j)\frac{\xi}{2\pi}}(\theta)\,,
\hspace{.5cm}n\,\,\,\mbox{odd} \\ \\
{\cal P}_\frac12(\theta)
\prod_{j=1}^{(n-2)/2}{\cal W}_{\frac12+(n-2j)\frac{\xi}{2\pi}}(\theta)\,,
\hspace{.5cm}n\,\,\,\mbox{even} \\ 
\end{array}
\right.
\EN
and $P^{\Phi}_{A_jB_n}(\theta)$ are polynomials in $\cosh\theta$ to be fixed
through the conditions on the poles. In particular we have
\bea
&& F_{A_jB_1}^{\sigma_{j+1\,(mod\,2)}}(\theta)=
  -F_{B_1A_j}^{\sigma_{j+1\,(mod\,2)}}(\theta)=(-1)^j\,a\,
\frac{T_{(1+\xi/\pi)/2}(\theta)}{{\cal P}_{(1+\xi/\pi)/2}(\theta)}\\
&& F_{A_jB_2}^{\sigma_{j}}(\theta)=
   F_{B_2A_j}^{\sigma_{j}}(\theta)=(b\cosh\theta+c)\,\frac{T_{1/2}(\theta)
T_{1/2+\xi/\pi}(\theta)}{{\cal P}_{1/2}(\theta)
{\cal P}_{1/2+\xi/\pi}(\theta)}
\label{sigmaeven}
\eea
with $a$, $b$ and $c$ determined by the equations
\bea
&& \mbox{Res}_{\theta=i(\pi+\xi)/2}F^{\sigma_2}_{A_1B_1}(\theta)=
i\Gamma^{A_2}_{A_1B_1}F^{\sigma_2}_{A_2}\\
&& \mbox{Res}_{\theta=i(\pi/2+\xi)}F^{\sigma_2}_{A_2B_2}(\theta)=
i\Gamma^{A_2}_{A_2B_2}F^{\sigma_2}_{A_2}\\
&& \mbox{Res}_{\theta=i\pi/2}F^{\sigma_2}_{A_2B_2}(\theta)=
i\Gamma^{A_1}_{A_2B_1}\Gamma^{B_1}_{B_1B_2}F^{\sigma_2}_{A_1B_1}
(i(\pi-3\xi)/2)\,\,.
\eea
The last equation is the specialisation of (\ref{doubleff}) to the double
pole appearing at $\theta=i\pi/2$ in the $A_jB_{2k}$ scattering amplitudes 
and related to the diagram of Fig.~5 with $a=A_j$, $b=B_{2k}$, $d=e=B_k$ and
$c=A_i$, $i=j$ ($i\neq j$) for $k$ even (odd). It can be checked that the
matrix elements (\ref{sigmaeven}) determined in this way satisfy the 
asymptotic factorisation condition \cite{DSC}
\EQ
\lim_{\theta\rightarrow\infty}F^{\sigma_j}_{A_jB_{2k}}(\theta)=
F^{\sigma_j}_{A_j}\frac{F^{\mu_i}_{B_{2k}}}{\langle\mu\rangle}
\EN
with $k=1$. The condition
\EQ
\lim_{\theta\rightarrow\infty}\left|F^{\mu_1\pm\mu_2}_{A_jA_j}(\theta)\right|=
\frac{(F^{\sigma_j}_{A_j})^2}{\langle\mu\rangle}
\EN
fixes the relative normalisation between order and disorder operators.

\resection{Correlation functions}

Within the form factor approach, correlation functions are obtained through
the spectral sum
\bea
\langle\Phi_1(x)\Phi_2(0)\rangle &=& \sum_{n=0}^\infty\sum_{a_1,\ldots,a_n}
\int_{\theta_1>\ldots>\theta_n}
\frac{d\theta_1}{2\pi}\ldots\frac{d\theta_n}{2\pi} \nonumber \\
&& \langle 0|\Phi_1(0)|a_1(\theta_1)\ldots a_n(\theta_n)\rangle
\langle a_n(\theta_n)\ldots a_1(\theta_1)|\Phi_2(0)|0\rangle e^{-E_n|x|}\,,
\label{corr}
\eea
where 
\EQ
E_n=\sum_{k=1}^nm_{a_k}\cosh\theta_k
\EN 
denotes the total energy of the $n$-particle asymptotic state. 
\begin{figure}
\vspace{-1.cm}
\centerline{
\includegraphics[width=10cm,angle=-90]{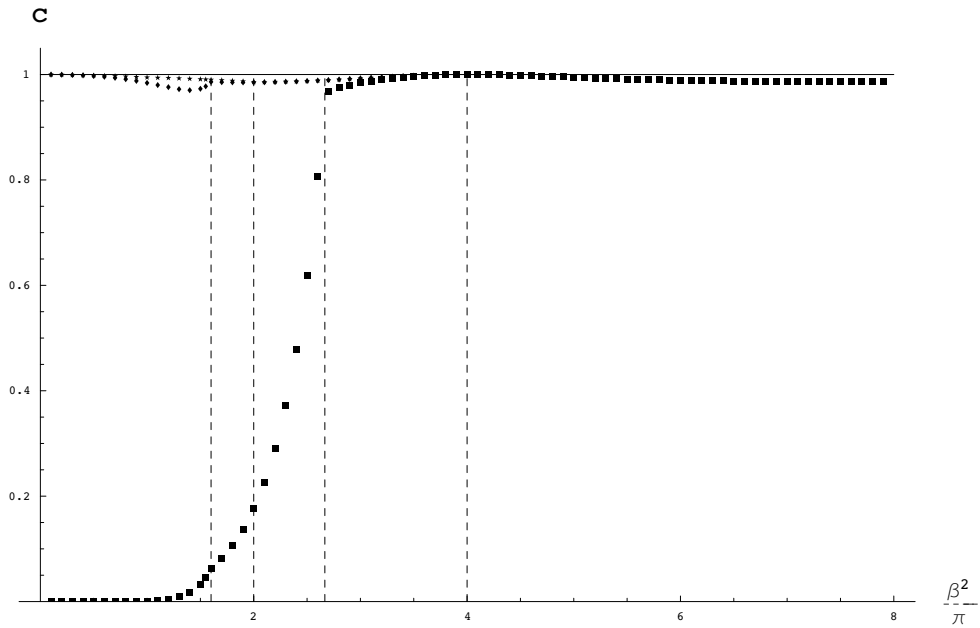}
}
\caption{\small{Different approximations of the central charge coming from
Eqs.~(\ref{cth}) and (\ref{corr}). The squares indicate the contribution
of the states $A_jA_j$ only, the diamonds the inclusion of the states $B_2$ 
and $B_1B_1$, the stars the inclusion of the states $B_4$, $B_1B_3$ and 
$B_2B_2$. The exact result is $c=1$.
From right to left, the dashed vertical lines correspond to the 
first four thresholds where the breather $B_n$ ($n=1,\ldots,4$) enters the 
spectrum of asymptotic particles.}}
\end{figure}
\begin{figure}
\centerline{
\includegraphics[width=10.cm,angle=-90]{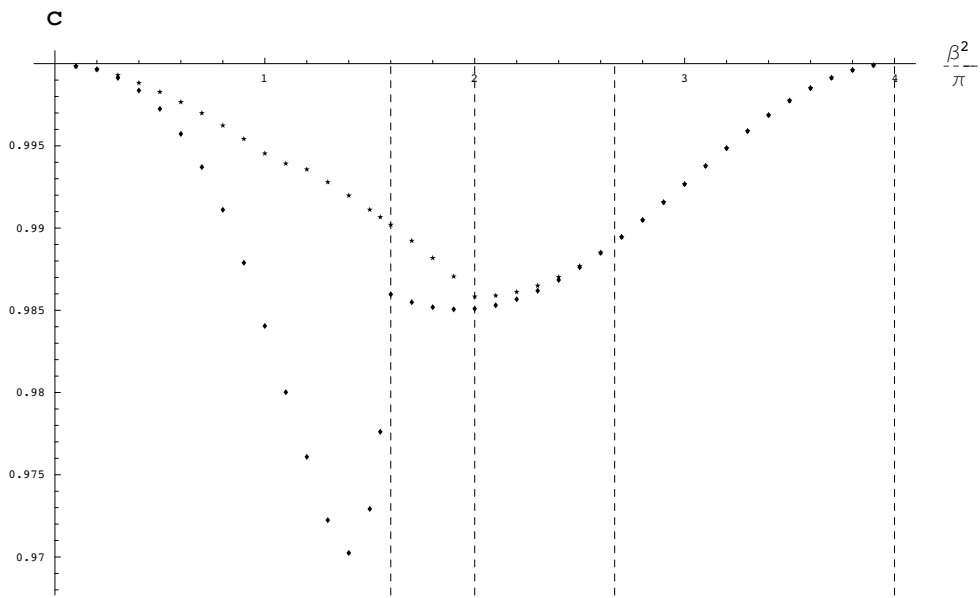}
}
\caption{\small{A detail of Fig.~6.}}
\end{figure}
This large distance expansion can produce good approximations of the 
integrated correlators even when only few lightest states are 
included in the sum. A quantitative illustration of the convergence pattern 
is obtained using the
spectral sum to compute exactly known quantities like the central charge $c$
and the scaling dimensions $X_\Phi$ through the sum rules \cite{cth,DSC}
\bea
c &=& \frac{3}{4\pi}\int d^2x\,|x|^2\langle\Theta(x)\Theta(0)\rangle_c
\label{cth}\\
X_\Phi &=& -\frac{1}{2\pi\langle\Phi\rangle}\int d^2x\,\langle\Theta(x)\Phi(0)
\rangle_c\,\,,
\label{x}
\eea
where $\langle\cdots\rangle_c$ denotes connected correlators and 
$\Theta(x)$ is the trace of the stress-energy tensor. The latter is 
proportional to the energy operator ${\cal E}(x)$ and its form factors are 
normalised through the condition 
\EQ
F^\Theta_{aa}(i\pi)=2\pi m_a^2\,\,.
\EN
Moreover, for $a\neq b$, $P^\Theta_{ab}(\theta)$ factorises
a term \cite{immf}
\EQ
\cosh\theta +\frac{m_a^2+m_b^2}{2m_am_b}\,\,.
\label{zero}
\EN
The asymptotic condition \cite{DSC}
\EQ
\lim_{\theta\rightarrow\infty}F^\Theta_{B_nB_n}(\theta)=
\frac{(F^\Theta_{B_n})^2}{\langle\Theta\rangle}
\EN
can be used to determine the expectation value
\bea
\label{vevth}
\langle\Theta\rangle=-\pi M^2\tan\frac\xi2  \,,
\eea
a result which coincides with that known from the thermodynamic Bethe ansatz
(see \cite{Alyosha}). 
\begin{figure}
\centerline{
\includegraphics[width=10cm,angle=-90]{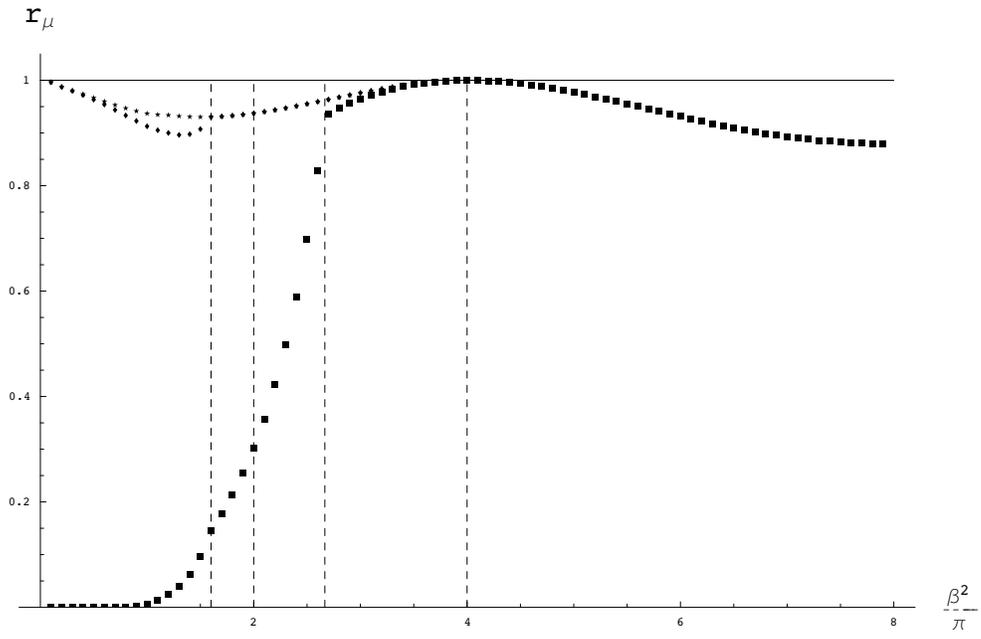}
}
\caption{The ratio (\ref{r}) for the 
operators $\mu_j$ coming from Eqs.~(\ref{x}) and (\ref{corr}).
The squares indicate the contribution
of the states $A_jA_j$ only, the diamonds the inclusion of the states $B_2$ 
and $B_1B_1$, the stars the inclusion of the state $B_4$. }
\end{figure}

\begin{figure}
\centerline{
\includegraphics[width=10cm,angle=-90]{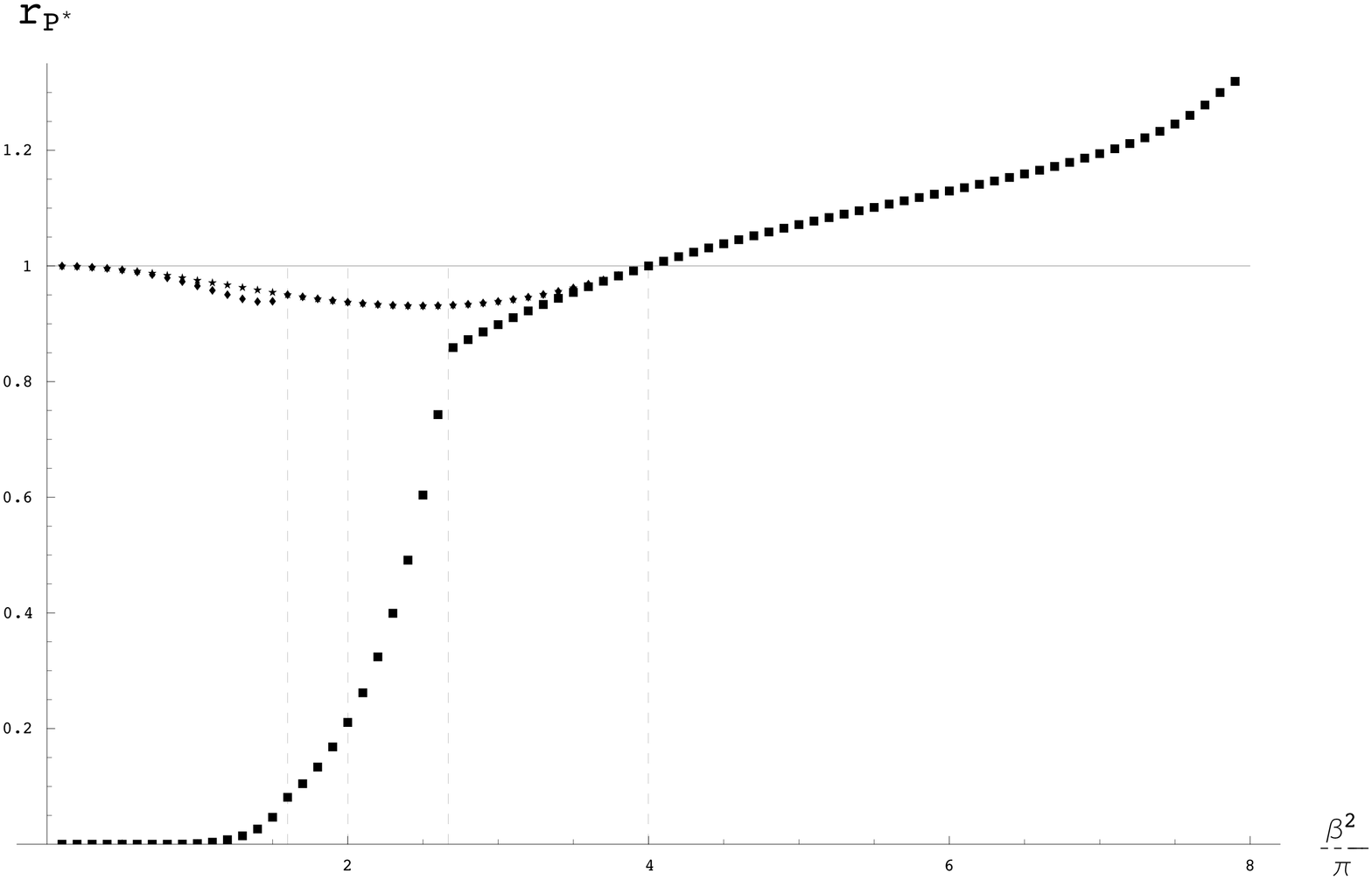}
}
\caption{As in Fig.~8 for the operator ${\cal P}^*$.}
\end{figure}

\begin{figure}
\centerline{
\includegraphics[width=10cm,angle=-90]{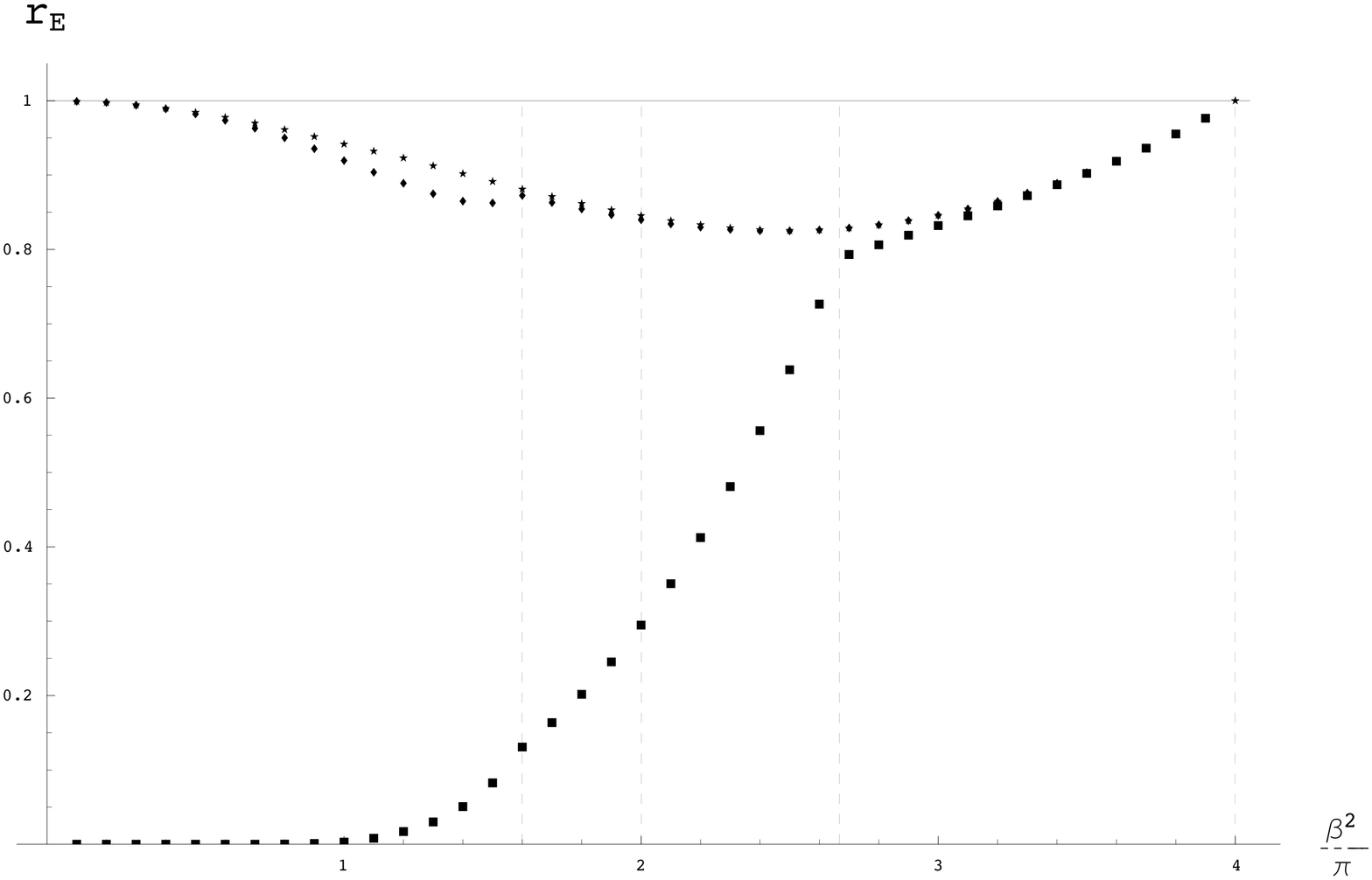}
}
\caption{As in Fig.~8 for the operator ${\cal E}$.}
\end{figure}
Figures 6 and 7 show the first few approximations provided by the insertion
of a truncated spectral sum into the formula (\ref{cth}) for the central 
charge. For this quantity as for the scaling dimensions computed through 
(\ref{x}), the states $A_jA_j$ yield the 
exact result at the free fermion point $\beta^2=4\pi$, as well as the state
$B_1B_1$ gives the exact result at the free boson point $\beta^2=0$. Away from
these free points the convergence of the series is extremely rapid due to the
factor $|x|^2$ in (\ref{cth}) which suppresses the contribution to the 
integral of the short distances, namely the region where the truncated 
spectral series fails to reproduce the exact correlator. Hence, in the 
repulsive region $4\pi<\beta^2<8\pi$ the $A_jA_j$ contribution reproduces the 
exact result with a maximal deviation which is slightly above $1\%$ as 
$\beta^2\rightarrow 8\pi$. Entering the attractive region below the free 
fermion point this contribution falls down quite rapidly as the particles 
$A_j$ become heavier than the lowest breathers (see (\ref{mn})). The inclusion
of the first few breather states, however, gives a quite accurate result
also in the repulsive region (Fig.~7). 

The same qualitative pattern can be observed in Figs.~8, 9 and 10 showing 
results for the ratios
\EQ
r_\Phi=\frac{X_\Phi^{approx}}{X_\Phi}
\label{r}
\EN
between the approximations for the scaling dimensions obtained inserting a
truncated spectral series in (\ref{x}) and the exact values. Quantitatively,
the absence 
of the ultraviolet suppressing factor in (\ref{x}) as compared to (\ref{cth})
obviously leads to poorer results for a given level of truncation. Of course,
the accuracy can be improved by including more states in the spectral sum. 
Here we simply observe that the increasing deviation from the exact value
for larger values of $\beta^2$ in the repulsive region is due to the 
increasing ultraviolet singularities of the exact correlators entering the 
sum rules, what makes increasingly important the contribution of the short
distances to the integrals. The result for $r_{\cal E}=r_\Theta$ is plotted 
up to the free fermion point because the corresponding integral in (\ref{x}) 
diverges for $\beta^2\geq 4\pi$ (see \cite{DSC}). The integral entering the 
computation of $r_{{\cal P}^*}$ becomes divergent at $\beta^2=8\pi$, what 
helps understanding why the two-particle approximation is particularly poor 
as this point is approached.

\resection{Universal ratios}
The ability to compute the correlation functions allows the evaluation of the
canonical thermodynamic observables. We will consider the magnetisations
\bea
&& M_\sigma=|\langle\sigma_j\rangle|\label{Msigma}\\
&& M_{\cal P}=|\langle{\cal P}\rangle|
\eea
the specific heat 
\EQ
C=\int d^2x\,\langle{\cal E}(x){\cal E}(0)\rangle_c
\EN
the susceptibilities
\bea
&& \chi_\sigma=\int d^2x\,\langle\sigma_j(x)\sigma_j(0)\rangle_c\\
&& \chi_{12}=\left|\int d^2x\,\langle\sigma_1(x)\sigma_2(0)\rangle_c
\right|\\
&& \chi_{\cal P}=\int d^2x\,\langle{\cal P}(x){\cal P}(0)\rangle_c
\eea
the second moment correlation lengths
\bea
&& \xi^{2nd}_\sigma=\left(\frac{1}{4\chi_\sigma}\int d^2x\,|x|^2
\langle\sigma_j(x)\sigma_j(0)\rangle_c\right)^{1/2}\\
&& \xi^{2nd}_{\cal P}=\left(\frac{1}{4\chi_{\cal P}}\int d^2x\,|x|^2
\langle{\cal P}(x){\cal P}(0)\rangle_c\right)^{1/2}
\eea
and the exponential correlation lengths $\xi_\Phi$ defined as
\bea
&& \lim_{|x|\rightarrow\infty}\langle\sigma_j(x)\sigma_j(0)\rangle_c\sim
e^{-|x|/\xi_\sigma}\\
&& \lim_{|x|\rightarrow\infty}\langle{\cal P}(x){\cal P}(0)\rangle_c\sim
e^{-|x|/\xi_{\cal P}}\,\,.\label{xip}
\eea

These quantities can be evaluated on the two sides of the critical line 
with continously varying exponents. In the disordered phase II we decompose 
the above correlators onto the form factors of section~5. Duality is 
exploited to obtain the results for the ferromagnetic phase I: the operators
are replaced by their duals and the correlators are again decomposed over 
the form factors of section~5; the observables above do not depend on which
of the four ferromagnetic ground states is selected by spontaneous symmetry 
breaking.

For a given observable ${\cal F}$, we denote by ${\cal F}^{+}$
its limit toward a given point on the critical line along a path in phase
II, and by ${\cal F}^-$ the limit toward the same point along the dual path
in phase I. Dimensionless numbers independent on metric factors can be obtained
suitably combining the limits toward the same fixed point of different 
observables. These numbers are universal and characterise the scaling region
around the critical line with continously varying exponents.

Some of these universal combinations can be determined exactly. It follows
from the spectral decomposition (\ref{corr}) that the exponential
correlation length $\xi_\Phi$ is simply the inverse mass of the lightest 
asymptotic state coupling to the operator $\Phi$. In the disordered phase
the lightest state is $A_j$ for $\sigma_j$; for ${\cal P}$ we have instead
$A_1A_2$ in the repulsive region $\beta^2>4\pi$ (i.e. $\xi>\pi$) and $B_1$ 
in the attractive region.
In the ferromagnetic phase, both $\sigma_j$ and ${\cal P}$ (or equivalently
$\mu_j$ and ${\cal P}^*$ in the disordered phase) couple to $A_jA_j$
for $\beta^2>8\pi/3$ (i.e. $\xi>\pi/2$) and to $B_2$ below this threshold.
The following universal ratios then follow from (\ref{mn})
\EQ
\frac{\xi^+_\sigma}{\xi^-_\sigma}=
\left\{
\begin{array}{l}
2\,,\hspace{.5cm}\xi>\pi/2 \\ \\
2\sin\xi\,,\hspace{.5cm}\xi<\pi/2 \\
\end{array}
\right.
\EN
\EQ
\frac{\xi^+_{\cal P}}{\xi^-_{\cal P}}=
\left\{
\begin{array}{l}
1\,,\hspace{.5cm}\xi>\pi \\ \\
1/\sin\frac\xi2\,,\hspace{.5cm}\frac\pi2<\xi<\pi \\ \\
2\cos\frac\xi2\,,\hspace{.5cm}\xi<\frac\pi2 \\ 
\end{array}
\right.
\EN
\EQ
\frac{\xi^+_\sigma}{\xi^+_{\cal P}}=
\left\{
\begin{array}{l}
2\,,\hspace{.5cm}\xi>\pi \\ \\
2\sin\frac\xi2\,,\hspace{.5cm}\xi<\pi\,\,. \\
\end{array}
\right.
\EN

The fact that the energy operator ${\cal E}$ is odd under duality and the 
specific heat is bilinear in ${\cal E}$ implies
\EQ
C^+/C^-=1\,\,.
\EN
The combination
\EQ
C^+ \; (\xi_\sigma^+)^2
= -(1-\alpha)(2-\alpha) \frac{\langle \Theta \rangle}{4 \pi M^2}\,,
\EN
where $\alpha=1-\frac{\xi}{\pi}$ is the specific heat critical exponent and 
$\langle \Theta \rangle$ is given in (\ref{vevth}), is also exact.

\begin{table}
\vspace{-2cm}
\hspace{-0.8cm}
\begin{center}
\begin{tabular}{|l|l|l|l|l|l|l|l|l|l|}
\hline
$\beta^2/\pi$ & 
$\frac{\xi^{2nd\,\,+}_\sigma}{\xi_\sigma^+}$ &
$\frac{\xi^{2nd\,\,+}_\sigma}{\xi^{2nd\,\,-}_\sigma}$ & 
$\chi_\sigma^+/\chi_\sigma^-$ 
& $\chi_{12}^- / \chi_\sigma^-$ & 
$R_\sigma$ & 
$\frac{\xi^{2nd\,\,+}_{\cal P}}{\xi_\sigma^+}$ & 
$\frac{\xi^{2nd\,\,+}_{\cal P}}{\xi^{2nd\,\,-}_{\cal P}}$ & 
$ \chi^+_{\cal P} / \chi^-_{\cal P} $ & $R_{\cal P}$ \\ 
\hline
2.0&0.9905&1.964&11.59&0.9411&1.434&0.9905&1.964&11.59&1.434 \vspace{-0.10cm} \\           
2.1&0.9923&2.070&13.42&0.9212&1.501&0.9322&1.931&10.79&1.320 \vspace{-0.10cm} \\
2.2&0.9938&2.176&15.33&0.8970&1.563&0.8795&1.897&10.06&1.218 \vspace{-0.10cm} \\
2.3&0.9950&2.279&17.32&0.8685&1.619&0.8316&1.863&9.392&1.125 \vspace{-0.10cm} \\
2.4&0.9961&2.381&19.33&0.8353&1.670&0.7880&1.829&8.774&1.040 \vspace{-0.10cm} \\
2.5&0.9970&2.479&21.36&0.7976&1.717&0.7482&1.795&8.202&0.9626 \vspace{-0.10cm} \\
2.6&0.9978&2.573&23.35&0.7555&1.758&0.7117&1.760&7.670&0.8915 \vspace{-0.10cm} \\
8/3&0.9983&2.634&24.64&0.7250&1.784&0.6891&1.737&7.334&0.8473 \vspace{-0.10cm} \\
2.7&0.9984&2.662&25.29&0.7093&1.796&0.6781&1.725&7.175&0.8264 \vspace{-0.10cm} \\
2.8&0.9988&2.745&27.15&0.6595&1.830&0.6470&1.689&6.720&0.7672 \vspace{-0.10cm} \\
2.9&0.9991&2.821&28.89&0.6065&1.861&0.6181&1.654&6.295&0.7127 \vspace{-0.10cm} \\
3.0&0.9993&2.890&30.51&0.5511&1.888&0.5915&1.618&5.897&0.6623 \vspace{-0.10cm} \\
3.1&0.9995&2.950&31.96&0.4939&1.911&0.5668&1.583&5.525&0.6156 \vspace{-0.10cm} \\
3.2&0.9997&3.002&33.26&0.4356&1.931&0.5438&1.548&5.175&0.5724 \vspace{-0.10cm} \\
3.3&0.9998&3.046&34.38&0.3769&1.948&0.5224&1.513&4.847&0.5323 \vspace{-0.10cm} \\
3.4&0.9999&3.081&35.33&0.3184&1.963&0.5026&1.479&4.538&0.4950 \vspace{-0.10cm} \\
3.5&0.9999&3.110&36.10&0.2608&1.975&0.4840&1.445&4.247&0.4602 \vspace{-0.10cm} \\
3.6&1.000&3.131&36.71&0.2045&1.984&0.4668&1.412&3.972&0.4278 \vspace{-0.10cm} \\
3.7&1.000&3.146&37.16&0.1500&1.991&0.4506&1.380&3.711&0.3976 \vspace{-0.10cm} \\
3.8&1.000&3.155&37.47&0.09763&1.996&0.4355&1.349&3.464&0.3694 \vspace{-0.10cm} \\
3.9&1.000&3.161&37.64&0.04756&1.999&0.4214&1.319&3.228&0.3430 \vspace{-0.10cm} \\
4.0&1&3.162&37.70&0&2&0.4082&1.291&3&0.3183 \vspace{-0.10cm} \\
4.1&1.000&3.161&37.65&0.04494&1.999&0.3959&1.264&2.782&0.2952 \vspace{-0.10cm} \\
4.2&1.000&3.157&37.50&0.08721&1.996&0.3843&1.238&2.579&0.2736 \vspace{-0.10cm} \\
4.3&1.000&3.152&37.28&0.1268&1.992&0.3735&1.213&2.389&0.2533 \vspace{-0.10cm} \\
4.4&1.000&3.145&36.98&0.1637&1.986&0.3634&1.189&2.211&0.2343 \vspace{-0.10cm} \\
4.5&1.000&3.138&36.63&0.1981&1.979&0.3539&1.165&2.044&0.2165 \vspace{-0.10cm} \\
4.6&1.000&3.130&36.23&0.2299&1.971&0.3450&1.142&1.887&0.1998 \vspace{-0.10cm} \\
4.7&1.000&3.122&35.78&0.2594&1.961&0.3366&1.120&1.740&0.1841 \vspace{-0.10cm} \\
4.8&1.000&3.114&35.30&0.2866&1.950&0.3288&1.099&1.603&0.1694 \vspace{-0.10cm} \\
4.9&1.000&3.105&34.80&0.3117&1.937&0.3215&1.078&1.473&0.1556 \vspace{-0.10cm} \\
5.0&1.000&3.097&34.28&0.3348&1.924&0.3147&1.059&1.352&0.1427 \vspace{-0.10cm} \\
5.1&1.000&3.090&33.74&0.3560&1.909&0.3082&1.040&1.238&0.1306 \vspace{-0.10cm} \\
5.2&1.000&3.082&33.18&0.3754&1.893&0.3022&1.022&1.132&0.1193 \vspace{-0.10cm} \\
5.3&1.000&3.075&32.62&0.3932&1.877&0.2966&1.005&1.032&0.1087 \vspace{-0.10cm} \\
5.4&1.000&3.069&32.05&0.4095&1.859&0.2914&0.9894&0.9383&0.09876 \vspace{-0.10cm} \\
5.5&1.000&3.063&31.48&0.4244&1.840&0.2865&0.9742&0.8509&0.08950 \vspace{-0.10cm} \\
5.6&1.000&3.057&30.91&0.4380&1.821&0.2819&0.9599&0.7691&0.08086 \vspace{-0.10cm} \\
5.7&1.000&3.052&30.33&0.4503&1.800&0.2776&0.9463&0.6929&0.07280 \vspace{-0.10cm} \\
5.8&1.000&3.047&29.76&0.4616&1.778&0.2736&0.9336&0.6218&0.06530 \vspace{-0.10cm} \\
5.9&1.000&3.043&29.18&0.4719&1.756&0.2699&0.9216&0.5557&0.05834 \vspace{-0.10cm} \\
6.0&1.000&3.039&28.60&0.4812&1.732&0.2664&0.9104&0.4944&0.05188\\
\hline
\end{tabular}
\caption{Two-particle approximation for the universal ratios along the 
Ashkin-Teller critical line. The relation between $\beta$ and the values
of the lattice couplings $J$ and $J_4$ at the corresponding critical point
is provided by Eqs.~(\ref{betaJ}) and (\ref{selfdual}). The values
$\beta^2/\pi=2,8/3,4$ correspond to the $4$-state Potts, decoupled Ising and
Fateev-Zamolodchikov models, respectively.}
\end{center}
\end{table}
The universal ratios involving the susceptibilities and second moment 
correlation lengths cannot be computed exactly. In Table~4 we list
the results provided by the form factor approach including in the spectral 
series all the one and two-particle states (two-particle approximation). 
The quantities $R_\sigma$ and $R_{\cal P}$ are defined as
\EQ
R_{\Phi}=\frac{\chi_\Phi^+}{(\xi_\sigma^+ M_\Phi^-)^2}\,\,.
\EN

An 
important indication about the size of the error involved in the two-particle 
approximation comes from the comparison with the exact results \cite{McCoy} 
avalaible for
the point $\beta^2=4\pi$, where the system reduces to two decoupled Ising
models. We stress that, although the theory at this point
is free, the opearators $\sigma_j$ and ${\cal P}$ belong to the non--trivial
sector and have non--zero form factors on an arbitrary number of particles. 
Hence, the results obtained for their correlators are representative of 
what happens at generic values of $\beta$. Table~5 shows that the error
of the two--particle approximation is extremely small (less than $0.1\%$)
for the ratios involving $\sigma_j$ only, while it grows to order $1\%$ 
for the ratios involving ${\cal P}$. This fact has a very clear origin.
The spectral representation we use for the correlators is a large distance 
expansion and when we truncate it to obtain approximated results we make 
an error on the `short' distances. Hence, the error on the integrals over 
all distance scales grows with the strength of the ultraviolet singularities
of the correlators, namely with the scaling dimensions of the operators.
The scaling dimension of ${\cal P}$ is twice that of $\sigma_j$ at 
$\beta^2=4\pi$ and this explains the two different error scales. 

On these grounds we expect that the error on the ratios involving only 
$\sigma_j$ will stay quite small along the whole critical line, as a 
consequence of the fact that the scaling dimension of this operator does not
depend on $\beta$. Concerning the ratios involving ${\cal P}$, the value
$X_{\cal P}=\frac{\beta^2}{16\pi}$ suggests that the error will be of the same size 
of that of the $\sigma$--ratios at $\beta^2=2\pi$ and then will increase 
with $\beta$ to values that (from the results of the previous section)
should not exceed $10\%$ at $\beta^2=6\pi$.

\begin{table}
\begin{center}
\begin{tabular}{|l|l|l|}\hline
Ratio & Two-particle & Exact\\
      & approximation &     \\ \hline
$ \xi_\sigma^{2nd\,+}/\xi_\sigma^+$ & $1$ & $0.999598087.. $ \\
$ \xi_\sigma^{2nd\,+}/\xi_\sigma^{2nd\,-}$ & $3.1623$ & $3.16249504.. $ \\
$ \chi_\sigma^+/\chi_\sigma^-$ & $37.699$ & $37.6936520.. $  \\
$ R_\sigma$ & $2$ & $2.00163051.. $  \\
$ \xi_{\cal P}^{2nd\,+}/\xi_\sigma^+$ & $0.408$ & $0.40656.. $ \\
$ \xi_{\cal P}^{2nd\,+}/\xi_{\cal P}^{2nd\,-}$ & $1.291$ & $1.3088.. $ \\
$ \chi_{\cal P}^+/\chi_{\cal P}^-$ & $3$ & $2.9108.. $ \\
$ R_{\cal P}$ & $0.318$ & $0.32104.. $  \\
\hline
\end{tabular}
\caption{Universal ratios at the Ising decoupling point
$\beta^2=4\pi$.}
\end{center}
\end{table}

For $\beta^2=2\pi$ the results we obtained should reproduce those for the 
$4$-state Potts model. In order to check this point we label
$\alpha=(\alpha_1,\alpha_2)$, with $\alpha_{1,2}=\pm 1$, the four states in
which each site of the lattice can be. Then we build out of the Ashkin--Teller
spin variables $\sigma_1$ and $\sigma_2$ the site variable 
$\sigma=(\sigma_1,\sigma_2)$ and introduce the traditional Potts spin
variables
\bea
s_\alpha(x) &=& \delta_{\sigma(x),\alpha}-\frac14\nonumber\\
&=& \frac14[\alpha_1\sigma_1(x)
+\alpha_2\sigma_2(x)+\alpha_1\alpha_2\sigma_1\sigma_2(x)]
\label{pottsAT}
\eea
satisfying $\sum_\alpha s_\alpha=0$. If we denote by $|0_\gamma\rangle$ the 
ground state that spontaneous symmetry breaking has selected in the 
ferromagnetic phase I, we will have
\bea
\langle 0_\gamma|s_\alpha(x)s_\alpha(0)|0_\gamma\rangle &=& \frac{1}{16}
\langle 0_\gamma|\sigma_1(x)\sigma_1(0)+\sigma_2(x)\sigma_2(0)+
\sigma_1\sigma_2(x)\sigma_1\sigma_2(0)+2\alpha_1\alpha_2\sigma_1(x)\sigma_2(0)
\nonumber\\
&+& 2\alpha_2\sigma_1(x)\sigma_1\sigma_2(0)+
2\alpha_1\sigma_2(x)\sigma_1\sigma_2(0)|0_\gamma\rangle\,\,.
\eea
Along the Potts trajectory the internal symmetry is enhanced to invariance
under global permutations of the four colours and one finds the expression 
\bea
&& \langle 0_{(1,1)}|s_\alpha(x)s_\alpha(0)|0_{(1,1)}\rangle_{Potts}=
\nonumber\\
&& \frac{1}{16}
\langle 0_{(1,1)}|3\sigma_1(x)\sigma_1(0)+2(\alpha_1\alpha_2+\alpha_2+\alpha_1)
\sigma_1(x)\sigma_2(0)|0_{(1,1)}\rangle_{Potts}= \nonumber\\
&& \frac{1}{16}
\langle 0_{(1,1)}|3\sigma_1(x)\sigma_1(0)+2(4\delta_{\alpha,{(1,1)}}-1)
\sigma_1(x)\sigma_2(0)|0_{(1,1)}\rangle_{Potts}\,\,
\label{pottsminus}
\eea
explicitely showing that only the cases $\alpha=\gamma$ and $\alpha\neq\gamma$ 
are distinguished in the correlator 
$\langle 0_\gamma|s_\alpha(x)s_\alpha(0)|0_\gamma\rangle$.
In the disordered phase mixed correlators vanish by symmetry and we have
\EQ
\langle s_\alpha(x)s_\alpha(0)\rangle_{Potts}= 
\frac{3}{16}\langle\sigma_1(x)\sigma_1(0)\rangle_{Potts}\,\,.
\label{pottsplus}
\EN

In the Potts model we define\footnote{We drop the subscript {\em Potts} on the
correlators below.} the longitudinal spontaneous magnetisation
\EQ
M=\langle 0_\alpha|s_\alpha|0_\alpha\rangle\,,
\EN
the high-temperature susceptibility
\EQ
\chi=\int d^2x\,\langle s_\alpha(x)s_\alpha(0)\rangle_c\,,
\EN
the low-temperature longitudinal and transverse susceptibilities
\bea
&& \chi_L=\int d^2x\,\langle 0_\alpha|s_\alpha(x)s_\alpha(0)|0_\alpha\rangle_c
\\
&& \chi_T=\int d^2x\,\langle 0_\alpha|s_\gamma(x)s_\gamma(0)|0_\alpha\rangle_c
\hspace{1cm}\alpha\neq\gamma
\eea
and the second moment and exponential correlation lengths $\xi^{2nd}$ and 
$\xi$ computed from 
the correlator $\langle s_\alpha(x)s_\alpha(0)\rangle$ at high temperature
and $\langle 0_\alpha|s_\alpha(x)s_\alpha(0)|0_\alpha\rangle$ at low 
temperature. \\
The relations between these quantities and the Ashkin-Teller observables 
follow from Eqs.~(\ref{pottsAT}), (\ref{pottsminus}) and (\ref{pottsplus}).
In particular one obtains
\EQ
\chi^+/\chi_L^-=\left[\frac{\chi_\sigma^+/\chi_\sigma^-}{1+2\chi_{12}^-/
\chi_\sigma^-}\right]_{\beta^2=2\pi}\simeq 4.02
\EN
\EQ
\chi^-_T/\chi_L^-=\left[\frac{1-\frac23\chi_{12}^-/\chi_\sigma^-}
{1+2\chi_{12}^-/\chi_\sigma^-}\right]_{\beta^2=2\pi}\simeq 0.129\,\,.
\EN
These results agree\footnote{There is a slight deviation from the values 
quoted in Refs.~\cite{DC,DBC} due to the fact that the contributions of 
the states $A_jB_2$ and $B_2B_2$ had been neglected in those works.} with 
those of Refs.~\cite{DC,DBC} where the amplitude ratios for the $q$-state 
Potts model were computed. This non-trivial check eliminates the doubt
raised in \cite{DBC} about the result of \cite{DC} for the ratio 
$\chi^+/\chi_L^-$ in the $q$-state Potts model. For $q=3$ full agreement 
with the theoretical prediction was found in the lattice studies of 
Refs.~\cite{SBB,EG}.

\resection{Conclusion}
We have computed the universal ratios along the Ashkin-Teller critical line
with continously varying exponents using the field theoretical description
of the scaling limit provided by the sine-Gordon model. As discussed in the
Introduction, these results can be tested through numerical simulation or
series expansions on the lattice model. Up to now, lattice results for the
universal ratios along the Ashkin-Teller critical line exist only for the 
particular cases of the Ising decoupling point (where the exact values are  
known, see Table~5) and of the 4-state Potts model \cite{CTV,EG}. 
In the latter case the lattice results are compatible with the field 
theoretical predictions but are affected by large uncertainties originated 
by logarthmic corrections to scaling \cite{CNS,SS} coming from the marginal 
operator that is responsible for the end of critical line at the Potts point. 
Reducing the error bars at this point as well as obtaining estimates at other 
points along the Ashkin-Teller critical line appear as challanging tasks for 
future lattice studies.

\vspace{.6cm}
\noindent
{\bf Acknowledgements.} This work was partially supported by the European
Commission TMR programme HPRN-CT-2002-00325 (EUCLID). 
The work of P.G. is supported by the COFIN ``Teoria dei Campi, Meccanica
Statistica e Sistemi Elettronici''.


\newpage


\begin{thebibliography}{99}

\bibitem{PHA} V. Privman, P.C. Hohenberg and A. Aharony, Universal critical
point amplitude relations, in {\em Phase transition and critical phenomena},
Vol. 14, ed. C. Domb and J.L. Lebowitz (Academic Press, New York, 1991). 
\bibitem{BPZ} A.A. Belavin, A.M. Polyakov and A.B. Zamolodchikov, Nucl. Phys.
B 241 (1984) 333.
\bibitem{FQS} D. Friedan, Z. Qiu and S. Shenker, Phys. Rev. Lett. 52 (1984) 
1575.
\bibitem{AT} J. Ashkin and E. Teller, Phys. Rev. 64 (1943) 178.
\bibitem{Baxter} R.J. Baxter, {\em Exactly solved models in statistical
mechanics} (Academic Press, New York, 1982).
\bibitem{Wegner} F.J. Wegner, J. Phys. C 5 (1972) L131.
\bibitem{KB} L.P. Kadanoff and A.C. Brown, Ann. Phys. 121 (1979) 318.
\bibitem{Kadanoff} L.P. Kadanoff, Phys. Rev. B 22 (1980) 1405.
\bibitem{Karowski} M. Karowski, P. Weisz, Nucl. Phys. B 139 (1978) 445.
\bibitem{Smirnov} F.A. Smirnov, {\em Form Factors in Completely Integrable
Models of Quantum Field Theory} (World Scientific) 1992.
\bibitem{Lukyanov} S. Lukyanov, Mod. Phys. Lett. A 12 (1997) 2543.
\bibitem{LZ} S. Lukyanov and A.B. Zamolodchikov, Nucl. Phys. B 607 (2001) 437. 
\bibitem{Fan} C. Fan, Phys. Lett. A 39 (1972) 136.
\bibitem{FS} S.J. Ferreira and A.D. Sokal, Phys. Rev. B 51 (1995) 6727.
\bibitem{FZ} V.A. Fateev and A.B. Zamolodchikov, Sov. Phys. JETP 62 (1985) 215.
\bibitem{Ditzian} R.V. Ditzian, J.R. Banavar, G.S. Grest and L.P. Kadanoff,
Phys. Rev. B 22 (1980) 2542.
\bibitem{Coleman} S. Coleman, Phys. Rev. D 11 (1975) 2088.
\bibitem{Mandelstam} S. Mandelstam, Phys. Rev. D 11 (1975) 3026.
\bibitem{msg} G. Delfino and G. Mussardo, Nucl. Phys. B 516 (1998) 675.
\bibitem{q4} G. Delfino and J. Cardy, Phys. Lett. B 483 (2000) 303.
\bibitem{ZZ} A.B. Zamolodchikov and Al.B. Zamolodchikov, Ann.Phys. 120 (1979) 
253.
\bibitem{CT} S. Coleman and H.J. Thun, Comm. Math. Phys. 61 (1978) 31.
\bibitem{Goebel} C.J. Goebel, Prog. Theor. Phys. Suppl. 86 (1986) 261.
\bibitem{previous} G. Delfino, Phys. Lett. B 450 (1999) 196.
\bibitem{YZ} V.P. Yurov and Al.B. Zamolodchikov, Int. J. Mod. Phys. A6 (1991) 
3419. 
\bibitem{immf} G. Delfino, G. Mussardo, Nucl. Phys. B 455 (1995) 724.
\bibitem{cth} A.B. Zamolodchikov, JETP Lett. 43 (1986) 730.

J.L. Cardy, Phys. Rev. Lett. 60 (1988) 2709.
\bibitem{DSC} G. Delfino, P. Simonetti and J.L. Cardy, Phys. Lett. B 387 
(1996) 327.
\bibitem{Alyosha} Al.B. Zamolodchikov, Int. J. Mod. Phys. A10 (1995) 1125.
\bibitem{McCoy} T.T. Wu, B.M. McCoy, C.A. Tracy and E. Barouch, Phys. Rev.
B 13 (1978) 316.
\bibitem{DC} G. Delfino and J.L. Cardy, Nucl. Phys. B 519 (1998) 551.
\bibitem{DBC} G. Delfino, G.T. Barkema and J. Cardy, Nucl. Phys. B 565 (2000) 
521.
\bibitem{SBB} L. Shchur, P. Butera and B. Berche, Nucl. Phys. B 620 (2002) 579.
\bibitem{EG} I.G. Enting and A.J. Guttmann, Susceptibility amplitudes for the 
3- and 4-state Potts models, Physica A, in press.
\bibitem{CTV} M. Caselle, R. Tateo and S. Vinti, Nucl. Phys. B 562 (1999) 549.
\bibitem{CNS} J.L. Cardy, M. Nauenberg and D.J. Scalapino, Phys. Rev. B 22
(1980) 2560.
\bibitem{SS} J. Salas and A.D. Sokal, J. Stat. Phys. 88 (1997) 567.


\end{thebibliography}
\end{document}